\documentclass[11pt,preprint,aps,showkeys,onecolumn]{revtex4} 
\usepackage{amsmath}
\usepackage{amsfonts}
\usepackage{amssymb}
\usepackage{natbib}
\usepackage{graphicx}
\usepackage{epstopdf}
\usepackage{subcaption}
\usepackage{mathrsfs}
\usepackage{dcolumn}
\usepackage{bm}
\usepackage[section]{placeins}
\usepackage{hyperref}
\usepackage{xcolor}
\usepackage{tensor}
\begin{document}
\title{Phase Transition and Critical Phenomena of Charged Einstein-Maxwell-Scalar Black Holes}
\author{Zeming Zhuang$^{1}$}
\author{Kun Meng~$^{2}$}
\author{Hongsheng Zhang$^{1}$}
\email{sps\_zhanghs$@$ujn.edu.cn}
\affiliation{1.School of Physics and Technology, University of Jinan, 336 West Road of Nan Xinzhuang, Jinan, Shandong 250022, China}
\affiliation{2. School of Physics and Electronic Information, Weifang University, Weifang 261061, China}

\begin{abstract}
  We study the phase transition and critical phenomenon of charged black holes in Einstein-Maxwell-scalar (EMs) theory. Through comprehensive analysis of thermodynamic behaviors manifested in 
$P-V$ diagrams, $G(T,P)$ surfaces, and $C_P$ curves, we establish that these black holes exhibit van der Waals-type phase transition behavior. The derived critical exponents governing the phase transition show precise correspondence with both van der Waals gas-liquid systems, reinforcing the connection between black hole thermodynamics and mean field theory statistics. The findings reveal a crucial dependence of phase transition properties on the scalar charge parameter. A critical threshold emerges where phase transitions become prohibited when scalar charge exceeds a specific magnitude. However, the transition persists asymptotically as scalar charge approaches zero. The analysis further demonstrates nonlinear relationships between scalar charge and critical parameters: while small scalar charges induce increasing critical volume with charge magnitude, larger values produce an inverse trend. Critical temperature displays complementary behavior, maintaining monotonic variation under certain conditions while exhibiting inverse correlation with critical volume in others. Significantly, the transition points governing critical volume and temperature trends occur at distinct scalar charge values for different black holes, indicating a non-trivial parameter dependence. These results highlight the scalar charge's dual role as both an enabler and suppressor of phase transitions in EMs black holes, providing new insights into the interplay between geometric configurations and thermodynamic properties in modified gravity theories.

\end{abstract}
\maketitle

\section{Introduction}
Black hole thermodynamics, which integrates thermodynamics, statistical mechanics, general relativity and quantum mechanics, plays an important role in modern theoretical physics. Black hole thermodynamics was mainly established in the 1970s, and a key milestone was the seminal work by Hawking, Bekenstein, and others, who formulated the four laws of black hole thermodynamics similar to the four laws of thermodynamics\cite{Bardeen:1973gs, Bekenstein:1973ur, Hawking:1974rv, Hawking:1975vcx, Bekenstein:1972tm}. Nowadays, the study of black hole thermodynamics extends beyond the study of black holes themselves. Thermodynamic concepts such as gravitational entropy, phase transitions, and temperature have been widely applied to other gravitationally constrained systems where no black hole exists, and significant research progress has been made in these applications. These research findings suggest that there is a fundamental and essential connection between gravity and thermodynamics.

Because the black hole exhibits thermal effects, it can be treated as a thermodynamic system. This thermodynamic analogy allows us to analyze the phase transition of black holes. There are two main methods to study the black hole phase transition. The first approach examines the singular relationship between specific heat and isothermal compressibility, where the black hole mass $M$ is identified as the internal energy $U$ of the system, and the cosmological constant $\Lambda$ as a constant\cite{Hawking:1982dh}. The other is to study the van der Waals-type $P-V$ phase transition of asymptotic Anti-de Sitter (AdS) black holes in an extended phase space, where the mass $M$ of black holes and cosmological constant $\Lambda$ are regarded as the enthalpy $H$ and pressure $P$ of the system respectively\cite{Kastor:2009wy, Kubiznak:2012wp}. Studying phase transitions in gravitational systems may help to solve some extremely difficult problems outside of gravity. For example, the Hawking-Page phase transition was explained as the quark confinement/deconfinement transition\cite{Witten:1998zw} and the ``bald"/``hairy" RN-AdS black hole phase transition was explained as the normal/superconducting state transition in condensed matter physics\cite{Hartnoll:2008kx, Roberts:2008ns}.

In recent years, studies on the black hole phase transition have attracted significant attention from researchers. Research on phase transitions in gravitational systems has expanded its scope from the initial focusing on Schwarzschild-AdS black holes\cite{Hawking:1982dh} and RN-AdS black holes\cite{Kubiznak:2012wp} to Gauss-Bonnet-AdS black holes\cite{Cai:2001dz, Cai:2013qga}, Einstein-Horndeski black holes\cite{Hu:2018qsy}, FRW universes\cite{Cai:2005ra, Cai:2006rs, Abdusattar:2023hlj, Kong:2021dqd}, and other gravitational systems\cite{Cai:2007wz, Chu:2025zuz, Zou:2013owa, Du:2023heb, Ma:2016lwr, Li:2023sig, Zhang:2020odg, Du:2021xyc, Zhang:2022mce, Wei:2015iwa, Zeng:2015wtt, Ma:2019pya, Wei:2022dzw, Zhao:2022dgc, Gao:2021xtt}. The meaning of $P$ and $V$ in $P-V$ phase transition has also evolved from the initial mathematical correspondence to possessing a theoretical basis and physical significance\cite{Abdusattar:2021wfv}.

In the late stages of the evaporation of black holes caused by Hawking radiation (when the curvature approaches the Planck scale), general relativity breaks down and the quantum theory of gravity becomes necessary. String theory is a candidate theory for quantum gravity, and the study of its classical solutions (black holes/black strings) provides clues for understanding quantum gravity effects. For black holes with electric or magnetic charges (gauge U(1) field), string theory predictions differ from those of general relativity even before reaching the Planck scale. The reason for this discrepancy is the presence of a scalar field called the dilaton. Due to the presence of the dilaton scalar field, black holes in string theory fundamentally differ from those in classical general relativity in aspects such as singularity structure, extremal limit behavior, and thermodynamic properties\cite{Horowitz:1992jp, Lu:1995hm, Lu:2014maa, Ma:2022nwq, Ma:2023tcj}.

EMs theory is one of the most straightforward extensions of general relativity. The dilaton scalar field is minimally coupled to the gravitational field and non-minimally coupled to the kinetic term of the gauge U(1) field. The EMs gravity theory originally emerged as the bosonic sector of the $N=4$ supergravity theory with SU(4) and also appears as an effective field theory in the low-energy limit of heterotic string theory. In the EMs theory, the black hole no-hair theorem and the cosmic censorship hypothesis are also challenged\cite{Huang:2019lsl, Lu:2019jus, Liu:2015tqa, Lu:2013eoa, Lu:1996bd, Richarte:2021fbi, Cremmer:1977tt, Ferrara:1995ih, Yu:2020rqi, Yu:2018eqq, Gao:2004tu, Gao:2004tv}. This paper primarily investigates the phase transition of EMs black holes, which will to some extent reflect the statistical properties and internal structure of black holes.

The purpose of this paper is to study the $P-V$ phase transition of static charged black holes in EMs theory in extended phase space. The metric of EMs theory is asymptotic to AdS spacetime in planar coordinates with the cosmological constant $\Lambda=-3g^2$, where $P$ is the thermodynamic pressure and $V$ denotes its conjugate volume. The solution to EMs black holes used in this paper is given in the reference\cite{Huang:2019lsl}.

The organization of this paper is as follows. In section \ref{sec:two}, to ensure systematic exposition of theoretical constructs and establish a coherent notational framework, we show foundational results from key prior related works. Section \ref{sec:three} is dedicated to investigate the $P-V$ phase transition for static charged EMs black holes. In section \ref{sec:four}, we analyze the critical phenomena of the phase transition. Section \ref{sec:five} concludes this paper.

\section{Phase Transition of RN-AdS Black Holes and solutions in EMs theory} \label{sec:two}
In all subsections of this section, as the basis for the following research, we review the $P-V$ phase transition of RN-AdS black holes\cite{Kubiznak:2012wp} and the structure of the static charged EMs black hole solution\cite{Huang:2019lsl} respectively.

\subsection{$P-V$ Phase Transition of RN-AdS Black Holes} \label{subsec:two-one}
In order to review the study process of the RN-AdS black hole $P-V$ phase transition, we start from the metric and fundamental thermodynamic quantities to obtain the equation of state and the Gibbs free energy of RN-AdS black holes. In the Schwarzschild-like coordinate, the metric and the U(1) field of RN-AdS black holes read
\begin{equation}
\label{eq:1}
\text{d}s=-f\text{d}t^2+\frac{\text{d}r^2}{f}+r^2\text{d}\Omega_2^2 \, ,
\end{equation}
\begin{equation}
\label{eq:2}
F=\text{d}A \, , \qquad A=-\frac Qr \text{d}t \, ,
\end{equation}
where the component $f$ of the metric is given by
\begin{equation}
\label{eq:3}
f=1- \frac{2M}{r} + \frac{Q^2}{r^2} + \frac{r^2}{l^2} \, .
\end{equation}
In the above equations, $\text{d}\Omega_2^2$ is the standard element on $S^2$, $Q$ represents the total charge, $M$ represents the ADM mass of the black hole and $l$ is the curvature of AdS spacetime.
The black hole event horizon is at the largest root $r_0$ of $f(r)=0$.

The temperature and entropy of the black hole can be determined by using the Euclidean trick
\begin{equation}
\label{eq:4}
T= \frac{V'(r_0)}{4 \pi} = \frac{2}{4 \pi r_0} \left( 1+ \frac{2 r_0^2}{l^2}- \frac{M}{r_0} \right) \, ,
\end{equation}
\begin{equation}
\label{eq:5}
S= \frac A4 \, , \qquad A=4 \pi r_0^2 \,.
\end{equation}
The thermodynamic pressure $P$ of the black hole and its corresponding thermodynamic volume $V$ are respectively
\begin{equation}
\label{eq:6}
P= - \frac{\Lambda}{8\pi}= \frac{3}{8\pi} \frac{1}{l^2} \, ,
\end{equation}
\begin{equation}
\label{eq:7}
V= \frac 43 \pi r_0^3 \, .
\end{equation}
Fixed charge $Q$, the equation of state of the RN-AdS black hole can be obtained combining equations \eqref{eq:4} and \eqref{eq:6}
\begin{equation}
\label{eq:8}
P= \frac{T}{2 r_0} - \frac{1}{8 \pi r_0^2} + \frac{Q^2}{8 \pi r_0^4} \, , \quad r_0 = \left( \frac{3V}{4 \pi} \right) ^{1/3} \, .
\end{equation}

The geometric equation of state \eqref{eq:8} is translated to a physical one by performing dimensional analysis. The physical pressure and temperature are given by
\begin{equation}
\label{eq:9}
\text{Press} = \frac{\hbar c}{l_P^2} \, , \quad \text{Temp} = \frac{\hbar c}{k}T \, , \quad l_P^2 = \frac{\hbar G_N}{c^3} \, ,
\end{equation}
where $l_P^2$ represents the Planck length. Then the physical equation of state can be written as
\begin{equation}
\label{10}
\text{Press} = \frac{\hbar c}{l_P^2} P = \frac{k \text{Temp}}{2l_P^2 r_0}- \frac{\hbar c l_P^2}{8 \pi l_P^4 r_0^2} + \frac{\hbar c l_P^6 Q^2}{8 \pi l_P^8 r_0^4} \, .
\end{equation}
Comparing with the van der Waals equation
\begin{equation}
\label{eq:11}
\left( P + \frac{a}{v^2} \right) \left( v-b \right) = kT \, ,
\end{equation}
the specific volume $v$ should be identified with
\begin{equation}
\label{eq:12}
v = 2 l_P^2 r_0 \, .
\end{equation}
In other words it is the horizon radius $r_0$, rather than the thermodynamic volume $V$, that should be associated with the fluid volume.

In this study method, the mass $M$ of black holes is regarded as the enthalpy $H$ of the system. Then the Gibbs free energy $G$ of the system can be written as
\begin{equation}
\label{eq:13}
G \left( T, P \right) = H-TS = \frac14 \left( r_0 - \frac{8 \pi}{3} Pr_0^3 +\frac{3 Q^2}{r_0} \right) \, .
\end{equation}

According to the equation \eqref{eq:8}, the $P-V$ diagram of the RN-AdS black hole has a van der Waals-type phase transition. And according to the equation \eqref{eq:13}, the $G(P, T)$ diagram of the RN-AdS black hole has the swallowtail behavior that is a sign of the first-order phase transition. In addition, critical exponents of the RN-AdS black hole phase transition are the same as that of the van der Waals phase transition.

\subsection{Solution of Static Charged EMs Black Hole} \label{subsec:two-two}
In this subsection, the review of the static charged EMs black hole will begin with the Lagrangian\cite{Huang:2019lsl}
\begin{equation}
\label{eq:14}
\mathcal{L}= \sqrt{-g} \left( R - \frac12 \left( \partial \phi \right)^2 - V \left( \phi \right) - \frac14 Z \left( \phi \right)^{-1} F^2 \right) \,.
\end{equation}
The scalar field with potential $V(\phi)$ is minimally coupled to gravity, but non-minimally coupled to the Maxwell field with a generic function $Z(\phi)$. The charged static solution is considered as
\begin{equation}
\label{eq:16}
\text{d}s^2 = -h(r) \text{d}t^2 + \frac{\text{d}r^2}{f(r)} + r^2 \text{d} \Omega_{(D-2),\,k}^2 \,,\quad \phi=\phi(r) \,,\quad A=\xi(r) \text{d}t \,.
\end{equation}
where d$\Omega_{(D-2),\,k}^2$ is a $(D-2)$-dimensional Euclidean-signatured Einstein metric with the Ricci tensor $R_{ij}=k(D-1)g_{ij}$. The parameter $k$ can take three non-trivial discrete values, $k=1,0,-1$, and the corresponding maximal symmetric metrics describe the round sphere, torus, and hyperboloid in $(D-2)$ dimensions. The scalar field $\phi$ is assumed to be
\begin{equation}
\label{eq:15}
\phi = 2k_0 \, \text{arcsinh} \left[ \left( \frac qr \right)^{\Delta} \right] \,,
\end{equation}
where $k_0$, $q$, and $\Delta$ are constants and $q$ is scalar charge. 

In D-dimensions, the function $Z(\phi)$ is given by
\begin{equation}
\label{eq:17}
\begin{split}
Z= & \frac12 \gamma_2 \left( D-2 \right) \left( D-3 \right) C^{\frac{2 \kappa}{D-2}} S^{-\frac{D-3}{\Delta}}
+C^{\frac{-2D+4\kappa+4}{D-2}} \left( \frac12 \left( D-2 \right) \left( D-3 \right) C^2 + \kappa \Delta S^2 \right) \\
& \times \left( \gamma_1-\gamma_2 S^{-\frac{D-3}{\Delta}} \tensor[_2]{F}{_1} \left[ -\frac{D-3}{2\Delta},\, \frac{\kappa}{D-2};\, 1-\frac{D-3}{2\Delta};\, -S^2 \right] \right),
\end{split}
\end{equation}
where $\kappa=\Delta k_0^2$, $C \equiv \cosh \left( \frac{\phi}{2k_0} \right)$, $S \equiv \sinh \left( \frac{\phi}{2k_0} \right)$. The scalar potential is given by
\begin{equation}
\label{eq:18}
\begin{split}
V=& -\frac12 C^{\frac{-2D+4\kappa+4}{D-2}} \left( 2\left( D-2 \right)\left( D-1 \right)C^2-4\kappa \Delta S^2 \right) \\
& \times \left( g^2+\tilde{V}-\alpha S^{\frac{D-1}{\Delta}} \tensor[_2]{F}{_1} \left[ -\frac{D-1}{2\Delta},\, \frac{\kappa}{D-2};\, \frac{D-1}{2\Delta}+1;\, -S^2 \right] \right) \\
& -\alpha \left( D-2 \right) \left( D-1 \right) C^{\frac{2\kappa}{D-2}} S^{\frac{D-1}{\Delta}} \,,
\end{split}
\end{equation}
\begin{equation}
\label{eq:19}
\begin{split}
\tilde{V}=\frac{k\bar{V}\left( \phi \right)}{q^2}-& \frac{k \left( D-2 \right) S^{\frac{2}{\Delta}} C^{\frac{2\left( D-2-2\Delta \right)}{D-2}}}{q^2\left( D-1 \right) \left( D-2 \right) C^2 - \kappa \Delta S^2} \\
& \times \left( 2C^{\frac{2\kappa}{D-2}} \tensor[_2]{F}{_1} \left[ -\frac{D-3}{2\Delta},\, \frac{\kappa}{D-2};\, \frac{D-3}{2\Delta}-1;\, -S^2 \right] +D-3 \right) \,,
\end{split}
\end{equation}
where the function $\bar{V}(\phi)$ can be expressed as a quadrature, given by
\begin{equation}
\label{eq:20}
\bar{V}\left( \phi \right)=\int{\text{d}\phi \frac{C^{1-\frac{2\kappa}{D-2}}S^{\frac{2}{\Delta}-1}}{\kappa} \tensor[_2]{F}{_1} \left[ -\frac{D-3}{2\Delta},\, \frac{\kappa}{D-2};\, \frac{D-3}{2\Delta}-1;\, -S^2 \right]} \,.
\end{equation}
The corresponding D-dimensional static charged black hole solution can be obtained by simultaneously solving Einstein field equation, scalar field equation and Maxwell field equation.
\begin{equation}
\label{eq:21}
\begin{aligned}
&\begin{split}
h=& g^2 r^2 + kr^2 \hat{h} - \frac{\alpha q^{D-1}}{r^{D-3}}\tensor[_2]{F}{_1} \left[ \frac{D-1}{2\Delta},\, \frac{\Delta k_0^2}{D-2};\, \frac{D+2\Delta-1}{2\Delta};\, -\left( \frac qr \right)^{2\Delta} \right] \\
&+\frac{\gamma_1 Q^2}{4r^{2\left( D-3 \right)}} - \frac{\gamma_2 Q^2}{4\left( rq \right)^{D-3}}\tensor[_2]{F}{_1} \left[-\frac{D-3}{2\Delta},\, \frac{\Delta k_0^2}{D-2};\, \frac{-D+2\Delta+3}{2\Delta};\, -\left( \frac qr \right)^{2\Delta} \right] \,,
\end{split}\\
&f=\sigma^2h \,,\quad
\sigma=\left( 1+\frac{q^{2\Delta}}{r^{2\Delta}} \right)^{\frac{k_0^2\Delta}{D-2}} \,,\quad
\phi=2k_0\text{arcsinh}\left( \frac qr \right)^{\Delta} \,,\\
&\xi=-\frac{\left( D-2 \right) \gamma_1 Q \sigma}{2r^{D-3}}+\frac{\left( D-2 \right) \gamma_2 Q}{2q^{D-3}}\left( \sigma \tensor[_2]{F}{_1} \left[-\frac{D-3}{2\Delta},\, \frac{\kappa}{D-2};\, 1-\frac{D-3}{2\Delta};\, \left( \frac qr \right)^{2\Delta} \right]-1 \right) \,,
\end{aligned}
\end{equation}
where
\begin{equation}
\label{eq:22}
\hat{h}'=-\frac{2}{r^3\sigma} \tensor[_2]{F}{_1} \left[-\frac{D-3}{2\Delta},\, \frac{\kappa}{D-2};\, \frac{D-3}{2\Delta}-1;\, -\left( \frac qr \right)^{2\Delta} \right] \,.
\end{equation}
The corresponding physical quantities mass $M$, temperature $T$, entropy $S$, electric charge $Q_e$, electric potential $\Phi_e$, pressure $P$, and thermodynamic volume $V_{th}$ are given by
\begin{equation}
\label{eq:23}
\begin{split}
& M=\frac{\left( D-2 \right)\omega}{64\pi} \left( 4\alpha q^{D-1}+\gamma_2 q^{3-D}Q^2 \right) \,,\\
& T=\frac{h'\left( r_0 \right) \sigma \left( r_0 \right)}{4\pi}\,,\quad S=\frac14 \omega r_0^{D-2}\,,\\
& Q_e =\frac{1}{16\pi}\omega Q \,,\quad \Phi_e = -\xi \left( r_0 \right) \,,\\
& P=\frac{1}{16\pi} \left( D-1 \right) \left( D-2 \right)g^2 \,,\quad V_{\text{th}}=\frac{\omega}{D-1}r_0^{D-1} \sigma \left( r_0 \right) \,,
\end{split}
\end{equation}
where $\omega=\int{\text{d}\Omega_{\left( D-2 \right),k}}$ stand for the volume of the codimension-two subspace.

We can prove that the thermodynamic quantities given by equation\eqref{eq:23} satisfy the first law of thermodynamics. Considering the scalar charge $q$ as a parameter of the theory, we first determine the total differentials of the metric $h$ and  the mass $M$ as follows
\begin{equation}
\label{eq:add1}
\begin{split}
& \text{d}h=\frac{\partial h}{\partial r_0} \text{d}r + \frac{\partial h}{\partial P} \text{d}P + \frac{\partial h}{\partial Q} \text{d}Q \,,\\
& \text{d}M = \frac{\text{d}M}{\text{d}Q} \text{d}Q \,.
\end{split} 
\end{equation}
Considering $\text{d}h=0$ at the horizon of the black hole, the first law of thermodynamics can be derived from equation\eqref{eq:add1} as follows
\begin{equation}
\label{eq:add2}
\begin{split}
& \hphantom{\Rightarrow \,\,\,} \sigma \text{d}h - \text{d}M + \frac{\text{d}M}{\text{d}Q} \text{d}Q = T\text{d}S + V \text{d} P - \text{d}M + \Phi_e \text{d} Q_e = 0 \\
& \Rightarrow \text{d}M = T\text{d}S + V \text{d} P + \Phi_e \text{d} Q_e \,.
\end{split}
\end{equation}

We study four cases in the reference\cite{Huang:2019lsl}.\\
\textbf{Case 1.} \, $D=4$, $\Delta=1$, $k_0=1$ :
\begin{equation}
\label{eq:24}
\begin{split}
& h=g^2 r^2+k+\frac 12 \alpha \left( -3q\sqrt{r^2+q^2}+3r^2 \text{arcsinh} \left( \frac qr \right) \right)+ \frac{ \left( \gamma_1 q-\gamma_2 \sqrt{r^2+q^2} \right) Q^2}{4qr^2} \,,\\
& f=\sigma^2h \,,\quad \sigma^2=1+\frac{q^2}{r^2} \,,\quad \xi=\frac{\left( \gamma_2q-\gamma_1\sqrt{r^2+q^2} \right)Q}{r^2} \,,\quad \phi=2 \text{arcsinh} \left( \frac qr \right).
\end{split}
\end{equation}
\textbf{Case 2.} \, $D=4$, $\Delta=1$, $k_0^2=3$ :
\begin{equation}
\label{eq:25}
\begin{split}
& \begin{split}
h=& g^2r^2+k\left( \frac{2r^2}{q^2} \log \left(1+\frac{q^2}{r^2} \right)-\frac{r^2}{r^2+q^2} \right)+3\alpha r^2 \left( \frac{q}{\sqrt{r^2+q^2}}-\text{arcsinh}\left( \frac qr \right) \right) \\
& +\frac{\left( \gamma_1q \sqrt{r^2+q^2}-\gamma_2 \left( r^2+2q^2 \right) \right)Q^2}{4qr^2 \sqrt{r^2+q^2}} \,,\quad f=\sigma^2 h \,,
\end{split}\\
& \begin{split}
\sigma^2=\left( 1+\frac{q^2}{r^2} \right)^3 \,,\quad \xi=\frac{\gamma_2qQ\left( 3r^2+2q^2 \right)}{r^4}-\frac{\gamma_1Q\left( r^2+q^2 \right)^{3/2}}{r^4} \,.
\end{split}
\end{split}
\end{equation}
\textbf{Case 3.} \, $D=4$, $\Delta=1$, $k_0^2=5$ :
\begin{equation}
\label{eq:26}
\begin{split}
& \begin{split}
h=& g^2r^2+k\left( 1-\frac{q^4\left( 10r^2+9q^2 \right)}{9\left( r^2+q^2 \right)^3} \right)-\frac{\alpha q^3r^2}{\left( r^2+q^2 \right)^{3/2}} \\
& +\left( \frac{\gamma_1}{4r^2}-\frac{\gamma_2 \left( 3r^4+12q^2r^2+8q^4 \right)}{12qr^2 \left( r^2+q^2 \right)^{3/2}} \right)Q^2 \,,\quad \sigma^2=\left( 1+\frac{q^2}{r^2} \right)^5 \,,
\end{split}\\
& \begin{split}
f=\sigma^2 h\,,\quad \xi=\frac{\gamma_2qQ\left( 8q^4+20q^2r^2+15r^4 \right)}{3r^6}-\frac{\gamma_1Q\left( r^2+q^2 \right)^{5/2}}{r^6} \,.
\end{split}
\end{split}
\end{equation}
\textbf{Case 4.} \, $D=5$, $\Delta=D-3=2$, $k_0^2=\frac{D-2}{2\left( D-3 \right)}=\frac 34$ :
\begin{equation}
\label{eq:27}
\begin{split}
& h=g^2r^2+k-2\alpha \left( \sqrt{r^4+q^4}-r^2 \right)+\frac{\left( \gamma_1q^2-\gamma_2\sqrt{r^4+q^4} \right)Q^2}{4q^2r^4} \,,\quad \sigma^2=\left( 1+\frac{q^4}{r^4} \right) \,,\\
& f=\sigma^2 h \,,\quad \xi=-\frac{3\gamma_1Q\sqrt{r^4+q^4}}{2r^4}+\frac{3\gamma_2Q\left( \sqrt{r^8-q^8}-r^4 \right)}{2q^2r^4} \,,\quad \phi=\sqrt{3}\text{arcsinh}\left( \frac qr \right)^2 \,.
\end{split}
\end{equation}

\section{$P-V$ Phase Transition of Static Charged EMs Black Hole} \label{sec:three}
In this section, we will discuss the van der Waals-type $P-V$ phase transition of the above four cases by drawing $P-V$ diagrams, $G(T,P)$ diagrams, and $C_P$ curves. In order to determine some of the parameters contained in the black hole solutions given in the previous section, we first investigate the degradation of the black hole solution \eqref{eq:21} as the scalar charge $q=0$ before studying the black hole $P-V$ phase transition.

\subsection{Degradation of D-dimensional Static Charged EMs Black Holes}
When $q=0$, we note that the black hole mass $M$ given by \eqref{eq:23} maybe diverge. For equation \eqref{eq:21}, the condition $h(r_0)=0$ should be satisfied at the black hole horizon. Then the relationship between $Q^2$ and $q$ can be given by
\begin{equation}
\label{eq:28}
\begin{split}
& Q^2=q^{D-3} \frac{4r_0^{D-3}\left( g^2r_0^{D-1}+kr_0^{D-1}\hat{h}-\alpha q^{D-1}\tensor[_2]{F}{_1} \left[-\frac{D-3}{2\Delta},\, \frac{\Delta k_0^2}{D-2};\, \frac{-D+2\Delta+3}{2\Delta};\, -\left( \frac qr \right)^{2\Delta} \right] \right)}{\gamma_2 r_0^{D-3}\tensor[_2]{F}{_1} \left[-\frac{D-3}{2\Delta},\, \frac{\Delta k_0^2}{D-2};\, \frac{-D+2\Delta+3}{2\Delta};\, -\left( \frac qr \right)^{2\Delta} \right]-\gamma_1 q^{D-3}} \,,\\
& \lim_{q\to 0} \frac{Q^2}{q^{D-3}}=\frac 00= \frac{4\left( g^2r_0^{D-1}+k r_0^{D-1}\left( \frac{1}{r_0^2}+Const \right) \right)}{\gamma_2} = \text{finite value} \,,
\end{split}
\end{equation}
where $Const$ is a constant of integration from $\hat{h}$. This ensures that the mass $M$ remains non-divergent.

When $q \to 0$, the last parameter of the hypergeometric function in the expression of $h$ in the equation \eqref{eq:21} tend to zero, that is $\left( \frac qr \right)^{2\Delta} \to 0$. Then the hypergeometric function tend to 1, that is $\tensor[_2]{F}{_1} \left[-\frac{D-3}{2\Delta},\, \frac{\Delta k_0^2}{D-2};\, \frac{-D+2\Delta+3}{2\Delta};\, 0 \right] \to 1$. The expression $h$ with $q \to 0$ can be obtained by taking the integral of equation \eqref{eq:22} into the expression $h$ of \eqref{eq:21} and connecting with the expression $M$ of \eqref{eq:23}
\begin{equation}
\label{eq:29}
h=k\left( 1+r^2 Const \right)-\frac{16 \pi}{\left( D-2 \right)\omega} \frac{M}{r^{D-3}}+\frac{\gamma_1 Q^2}{4r^{2 \left( D-3 \right)}}+g^2r^2 \,.
\end{equation}
Considering that the EMs static charged black hole reduces to the RN-AdS black hole in four dimension when the scalar charge tend to zero, that is, the equation \eqref{eq:29} becomes the equation \eqref{eq:3}, we can get $Const=0$, $k=1$, and $\gamma_1=4$. When $q=0$, the expression $h$ reduces to
\begin{equation}
\label{eq:30}
h=1-\frac{2M}{r}+g^2r^2.
\end{equation}
This essentially corresponds to the components of the Schwarzschild black hole metric.

\subsection{$P-V$ Phase Transition}
The $P-V$ phase transition is marked by the inflection point in the $P-V$ diagram, i.e.,
\begin{equation}
\label{eq:31}
\left. \frac{\partial P}{\partial v} \right|_{v=v_0} =0 \,,\quad
\left. \frac{\partial^2P}{\partial v^2} \right|_{v=v_0} =0 \,,\quad
\left. \frac{\partial^2P}{\partial v^2} \right|_{v=v_0-\delta}>0 \,,\quad
\left. \frac{\partial^2P}{\partial v^2} \right|_{v=v_0+\delta}<0 \,,
\end{equation}
where $\delta$ is a positive infinitesimal. If the thermodynamic system indeed has a $P-V$ phase transition, the Gibbs $G(T,P)$ diagram of the system should manifest a swallowtail characteristic and the isobaric heat capacity $C_P$ is divergent accompanied by finite isochoric heat capacity $C_V$.

\textbf{Case 1.}
We still take $r_0$ as the location of the black hole horizon. Then from equations \eqref{eq:23} and \eqref{eq:24}, combining the expressions of $T$, $h$, and $\sigma^2$ we yield
\begin{equation}
\label{eq:32}
\begin{split}
& Q^2= \frac{4qr_0^2 \left( g^2 r_0^2+k+\frac 12 \alpha \left( -3q \sqrt{r_0^2+q^2}+3r_0^2 \text{arcsinh} \left( \frac{q}{r_0} \right) \right) \right)}{\gamma_2 \sqrt{r_0^2+q^2}-\gamma_1 q} \,,\\
& \begin{split}
T=\frac{1}{4\pi} \left(
\vphantom{\left. \frac{\gamma_2 \left( g^2r_0^2+k+ \frac 12 \alpha \left( -3q \sqrt{r_0^2+q^2}+3r_0^2 \text{arcsinh} \frac{q}{r_0} \right) \right)}{\gamma_1q-\gamma_2\sqrt{r_0^2+q^2}} \right)} \right.
& 2g^2 \sqrt{r_0^2+q^2}+\frac 12 \alpha \left( -6q+6\sqrt{r_0^2+q^2} \text{arcsinh}\left( \frac{q}{r_0} \right) \right) \\
& +\left. \frac{2\sqrt{r_0^2+q^2}\left( g^2r_0^2+k+ \frac 12 \alpha \left( -3q\sqrt{r_0^2+q^2}+3r_0^2 \text{arcsinh} \frac{q}{r_0} \right) \right)}{r_0^2} \right. \\
& +\left. \frac{\gamma_2 \left( g^2r_0^2+k+ \frac 12 \alpha \left( -3q \sqrt{r_0^2+q^2}+3r_0^2 \text{arcsinh} \frac{q}{r_0} \right) \right)}{\gamma_1q-\gamma_2\sqrt{r_0^2+q^2}} \right) \,.
\end{split}
\end{split}
\end{equation}
From the equations \eqref{eq:23} and \eqref{eq:32}, we obtain
\begin{equation}
\label{eq:33}
\begin{split}
P=\frac{3}{8\pi} \left( \vphantom{\left. \frac{\gamma_2 \left( g^2r_0^2+k+ \frac 12 \alpha \left( -3q \sqrt{r_0^2+q^2}+3r_0^2 \text{arcsinh} \frac{q}{r_0} \right) \right)}{\gamma_1q-\gamma_2\sqrt{r_0^2+q^2}} \right)} \right.
& 4\pi T-\frac{2\sqrt{r_0^2+q^2}\left( k+ \frac 12 \alpha \left( -3q\sqrt{r_0^2+q^2}+3r_0^2 \text{arcsinh} \frac{q}{r_0} \right) \right)}{r_0^2} \\
&-\frac{\gamma_2 \left( k+ \frac 12 \alpha \left( -3q \sqrt{r_0^2+q^2}+3r_0^2 \text{arcsinh} \frac{q}{r_0} \right) \right)}{\gamma_1q-\gamma_2\sqrt{r_0^2+q^2}} \\
& \left. \vphantom{\left. \frac{\gamma_2 \left( g^2r_0^2+k+ \frac 12 \alpha \left( -3q \sqrt{r_0^2+q^2}+3r_0^2 \text{arcsinh} \frac{q}{r_0} \right) \right)}{\gamma_1q-\gamma_2\sqrt{r_0^2+q^2}} \right)}
-\frac 12 \alpha \left( -6q+6\sqrt{r_0^2+q^2} \text{arcsinh}\left( \frac{q}{r_0} \right) \right) \right) \div \left( 4\sqrt{r_0^2+q^2}+\frac{\gamma_2 r_0^2}{\gamma_1q-\gamma_2 \sqrt{r_0^2+q^2}}  \right) \,.
\end{split}
\end{equation}
Considering $q << r_0$ and physical dimension, the trems involving $T$ in the expression for $P$ from equation \eqref{eq:33} can be approximated as
\begin{equation}
\label{eq:34}
\begin{split}
\text{Press} \approx \frac{3 k \text{Temp}}{2l_0^2 \left( 4r_0 -1 \right)}+\cdots \,.
\end{split}
\end{equation}
Comparing with the van der Waals equation \eqref{eq:11}, the specific volume $v$ can be identified with
\begin{equation}
\label{eq:35}
v=\frac 23 l_0^2 \left( 4r_0-1 \right).
\end{equation}
The horizon radius $r_0$ also is associated with the fluid volume. Combining equations \eqref{eq:23}, \eqref{eq:32}, and $h(r_0)=0$, the Gibbs free energy $G$ of the system is written as
\begin{equation}
\label{eq:36}
\begin{split}
G(T, P) & =H-TS=M-TS \\
& =\frac 12 \alpha q^3+\frac{\gamma_2 r_0^2 \left( g^2 r_0^2+k+ \frac 12 \alpha \left( -3q \sqrt{r_0^2+q^2}+3r_0^2 \text{arcsinh} \left( \frac{q}{r_0} \right) \right) \right)}{2\left( \gamma_2 \sqrt{r_0^2+q^2}-\gamma_1 q \right)} \\
&  \mathrel{\hphantom{=}}
\begin{split}
-\frac 14 \left(
\vphantom{\left. \frac{\gamma_2 \left( g^2r_0^2+k+ \frac 12 \alpha \left( -3q \sqrt{r_0^2+q^2}+3r_0^2 \text{arcsinh} \frac{q}{r_0} \right) \right)}{\gamma_1q-\gamma_2\sqrt{r_0^2+q^2}} \right)} \right.
& 2g^2 r_0^2 \sqrt{r_0^2+q^2}+\frac 12 \alpha r_0^2 \left( -6q+6\sqrt{r_0^2+q^2} \text{arcsinh}\left( \frac{q}{r_0} \right) \right) \\
& +2\sqrt{r_0^2+q^2}\left( g^2r_0^2+k+ \frac 12 \alpha \left( -3q\sqrt{r_0^2+q^2}+3r_0^2 \text{arcsinh} \frac{q}{r_0} \right) \right) \\
& +\left. \frac{\gamma_2  r_0^2\left( g^2r_0^2+k+ \frac 12 \alpha \left( -3q \sqrt{r_0^2+q^2}+3r_0^2 \text{arcsinh} \frac{q}{r_0} \right) \right)}{\gamma_1q-\gamma_2\sqrt{r_0^2+q^2}} \right) \,.
\end{split}
\end{split}
\end{equation}
In the equation \eqref{eq:37}, the isobaric heat capacity $C_P$ and the isochoric heat capacity $C_V$ are derived by coupling equations \eqref{eq:23} and \eqref{eq:32}.
\begin{equation}
\label{eq:37}
\begin{split}
& C_V=T \left. \frac{\partial S}{\partial T} \right|_V =0
\,,\\
& \begin{split}
C_P &=T \left. \frac{\partial S}{\partial T} \right|_P = T \left. \frac{\partial S}{\partial r_0} \right|_P \left/ \left. \frac{\partial T}{\partial r_0} \right|_P \right.\\
& =2 \pi  r_0^2 \sqrt{q^2+r_0^2} \left(\gamma _2 \sqrt{q^2+r_0^2}-\gamma _1 q\right) \left( \rule{0pt}{30pt} -2 \gamma _1 q \left( \rule{0pt}{25pt} -3 \alpha  q \left(q^2+2 r_0^2\right) \right. \right. \\
& \quad \left. \rule{0pt}{25pt} +2 \sqrt{q^2+r_0^2} \left(r_0^2 \left(2 g^2+3 \alpha  \sinh ^{-1}\left(\frac{q}{r_0}\right)\right)+k\right) \right) + \gamma _2 \left( \rule{0pt}{25pt} 2 k \left(2 q^2+r_0^2\right) \right. \\
& \quad \left.\rule{0pt}{30pt} \left. \rule{0pt}{25pt} r_0^2 \left(4 q^2+3 r_0^2\right) \left(2 g^2+3 \alpha  \sinh ^{-1}\left(\frac{q}{r_0}\right)\right)-3 \alpha  q \sqrt{q^2+r_0^2} \left(2 q^2+3 r_0^2\right) \right) \right) \left. \rule{0pt}{30pt} \right/ \\
& \quad \left( \rule{0pt}{30pt} 4 \gamma _1^2 q^2 \left(2 g^2 r_0^4-k \left(2 q^2+r_0^2\right)+3 \alpha  q \sqrt{q^2+r_0^2} \left(q^2-r_0^2\right)+3 \alpha  r_0^4 \sinh ^{-1}\left(\frac{q}{r_0}\right)\right) \right. \\
& \quad -2 \gamma _1 \gamma _2 q \left(\sqrt{q^2+r_0^2} \left(r_0^4 \left(6 g^2+9 \alpha  \sinh ^{-1}\left(\frac{q}{r_0}\right)\right)-8 k q^2-4 k r_0^2\right)+3 \alpha  q \left(4 q^4-3 r_0^4\right)\right) \\
& \quad \gamma _2^2 \left( \rule{0pt}{25pt} 4 g^2 q^2 r_0^4+6 g^2 r_0^6+3 \alpha  r_0^4 \left(2 q^2+3 r_0^2\right) \sinh ^{-1}\left(\frac{q}{r_0}\right) -2 k \left(4 q^4+6 q^2 r_0^2+r_0^4\right) \right. \\
& \quad \left.\rule{0pt}{30pt}\left. \rule{0pt}{25pt} +3 \alpha  q \sqrt{q^2+r_0^2} \left(4 q^4-3 r_0^4\right) \right) \right) \,.
\end{split}
\end{split}
\end{equation}
where the ``isochoric" of $C_V$ means that the thermodynamic volume $V_{th}$ in equation\eqref{eq:23} remains constant, i.e., $\text{d}V_{th}=0$. Since the scalar charge $q$ is a parameter in the theory, the $r_0$ must remain unchanged to maintain the condition $\text{d}V_{th}=0$. Therefore, $C_V=0$.

\begin{figure}[htbp]
\centering
\begin{subfigure}[b]{0.45\textwidth}
\centering
\includegraphics[width=\textwidth]{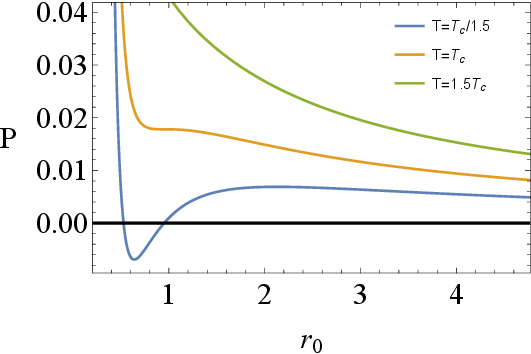}
\caption{}
\label{subfig:1PV}
\end{subfigure}
\hfill
\begin{subfigure}[b]{0.45\textwidth}
\centering
\includegraphics[width=\textwidth]{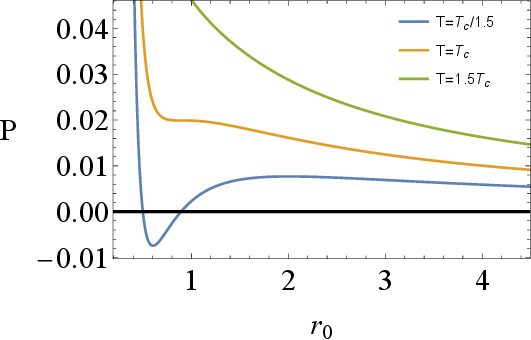}
\caption{}
\label{subfig:2PV}
\end{subfigure}

\begin{subfigure}[b]{0.45\textwidth}
\centering
\includegraphics[width=\textwidth]{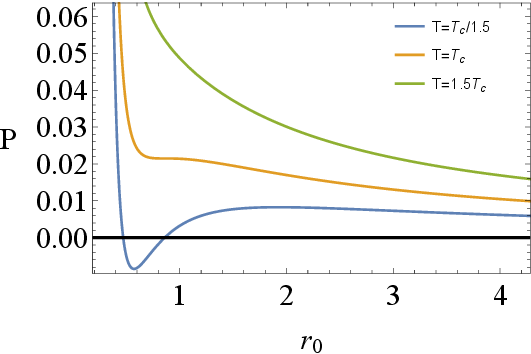}
\caption{}
\label{subfig:3PV}
\end{subfigure}
\hfill
\begin{subfigure}[b]{0.45\textwidth}
\centering
\includegraphics[width=\textwidth]{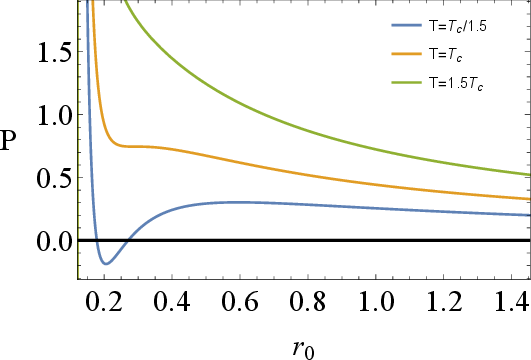}
\caption{}
\label{subfig:4PV}
\end{subfigure}
\caption{$P-V$ diagrams. (a), (b), (c) and (d) correspond to the Case 1, 2, 3 and 4 respectively. In these $P-V$ diagrams, the blue, orange and green curves from bottom to top correspond to the $P-V$ isothermals at $T=T_c/1.5,\,T=T_c$, and $T=1.5T_c$ respectively. These $P-V$ diagrams exhibit striking similarities to the characteristic of the van der Waals phase transition.}
\label{fig:PV}
\end{figure}

To analyze the $P-V$ phase transition, we first set the parameters in the Lagrangian \eqref{eq:14} as $\alpha=1$, $k=1$, $\gamma_1=4$, $\gamma_2=1$, and $q=0.05$. As constants in the Lagrangian, the parameters $\alpha$, $k$, $\gamma_1$, $\gamma_2$, and $q$ remain invariant in the four cases. Then the $P-V$ diagram corresponding to the equation \eqref{eq:33} is depicted in  FIG.\ref{subfig:1PV}. From FIG.\ref{subfig:1PV}, we find that the $P-V$ diagram has reminiscent behaviors with the van der Waals gas. In other words, the static charged EMs black hole has van der Waals-type phase transition when $D=4$, $\Delta=1$, and $k_0=1$. Substituting equation \eqref{eq:33} into the phase transition condition of equation \eqref{eq:31}, we obtain $P_c \approx 0.017787555$, $r_{0c} \approx 0.954774802$, and $T_c \approx 0.092728254$.

\begin{figure}[htbp]
\centering
\begin{subfigure}[b]{0.45\textwidth}
\centering
\includegraphics[width=\textwidth]{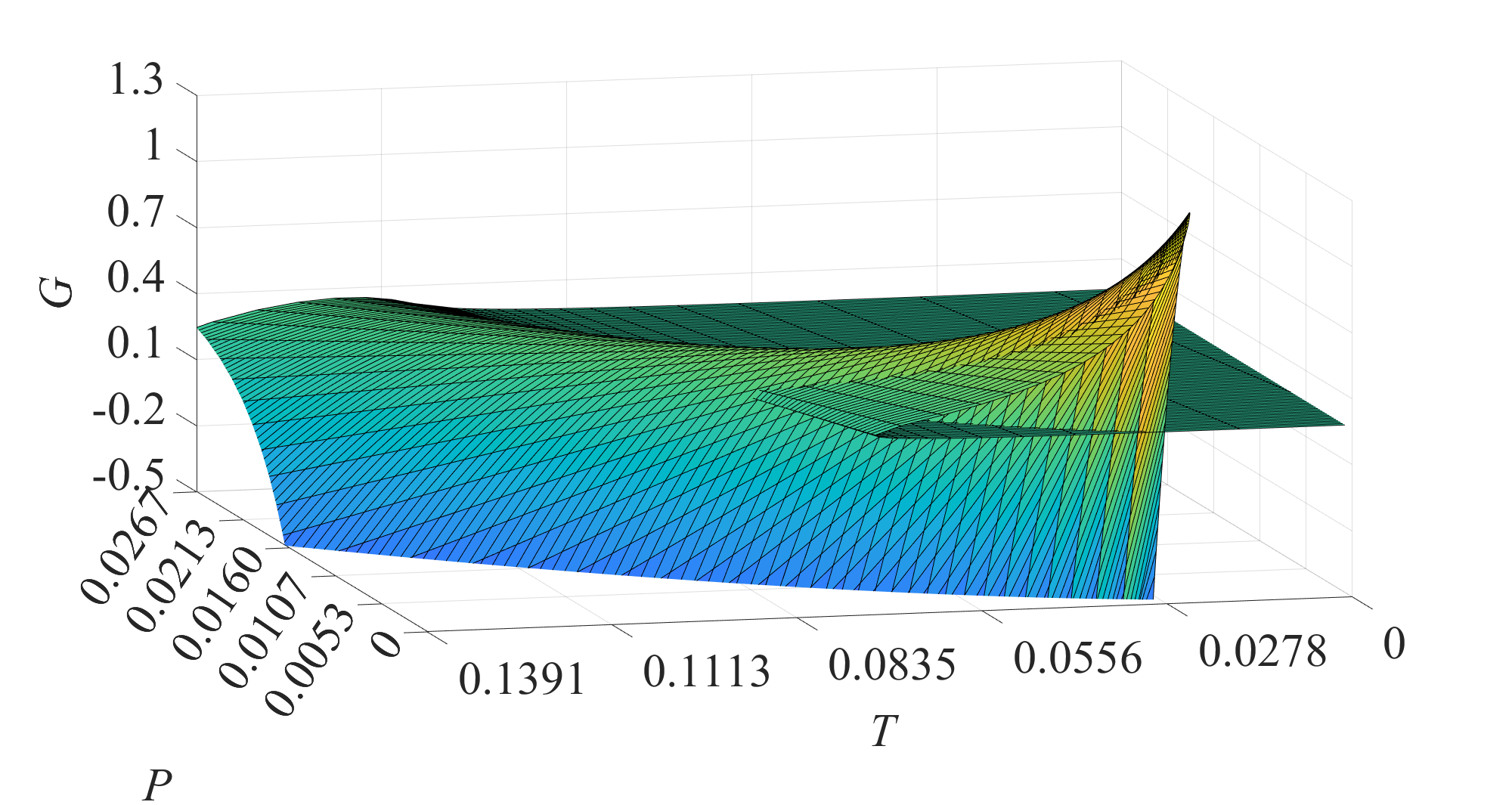}
\caption{}
\label{subfig:1Gibbs}
\end{subfigure}
\hfill
\begin{subfigure}[b]{0.45\textwidth}
\centering
\includegraphics[width=\textwidth]{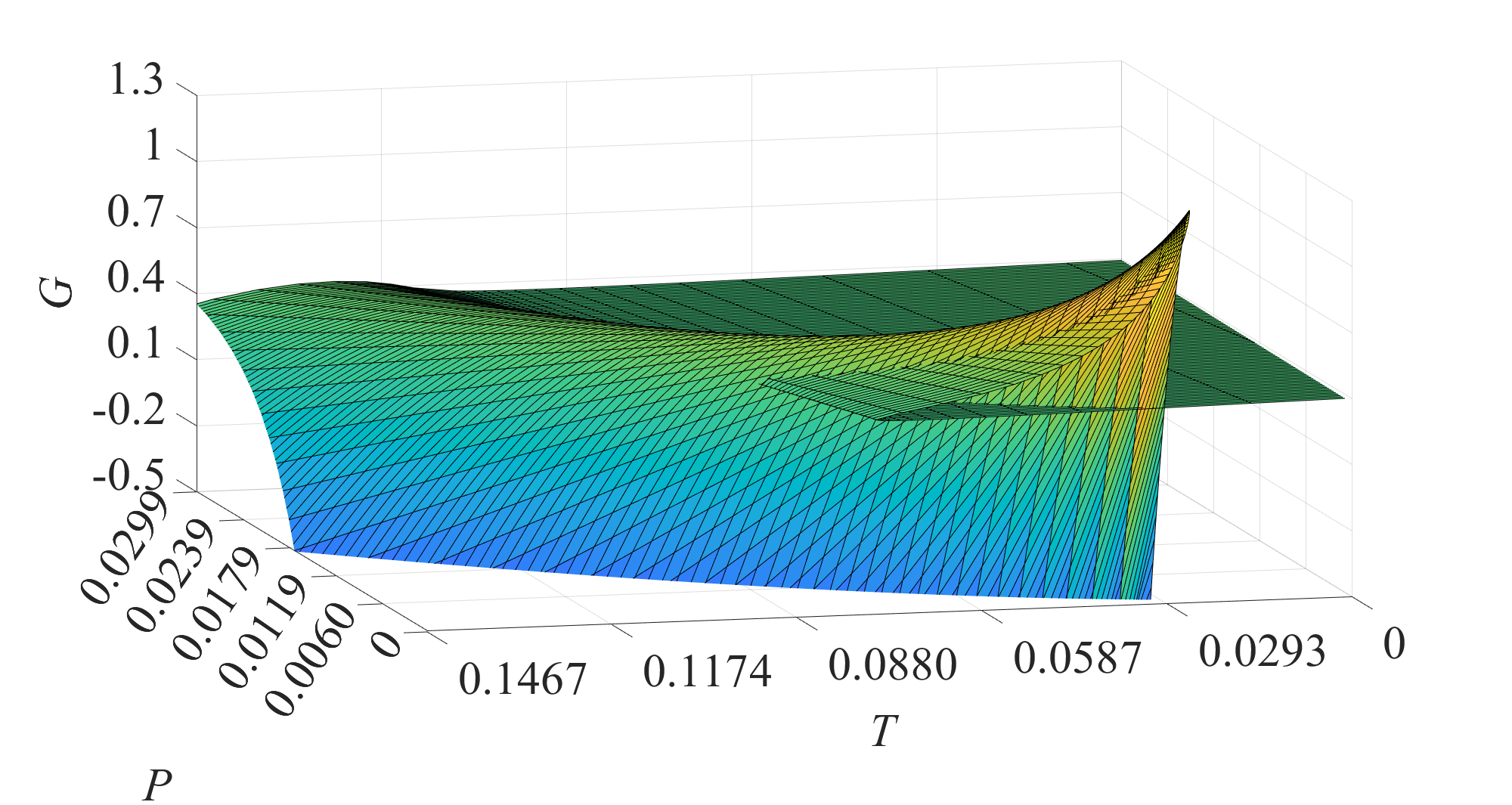}
\caption{}
\label{subfig:2Gibbs}
\end{subfigure}

\begin{subfigure}[b]{0.45\textwidth}
\centering
\includegraphics[width=\textwidth]{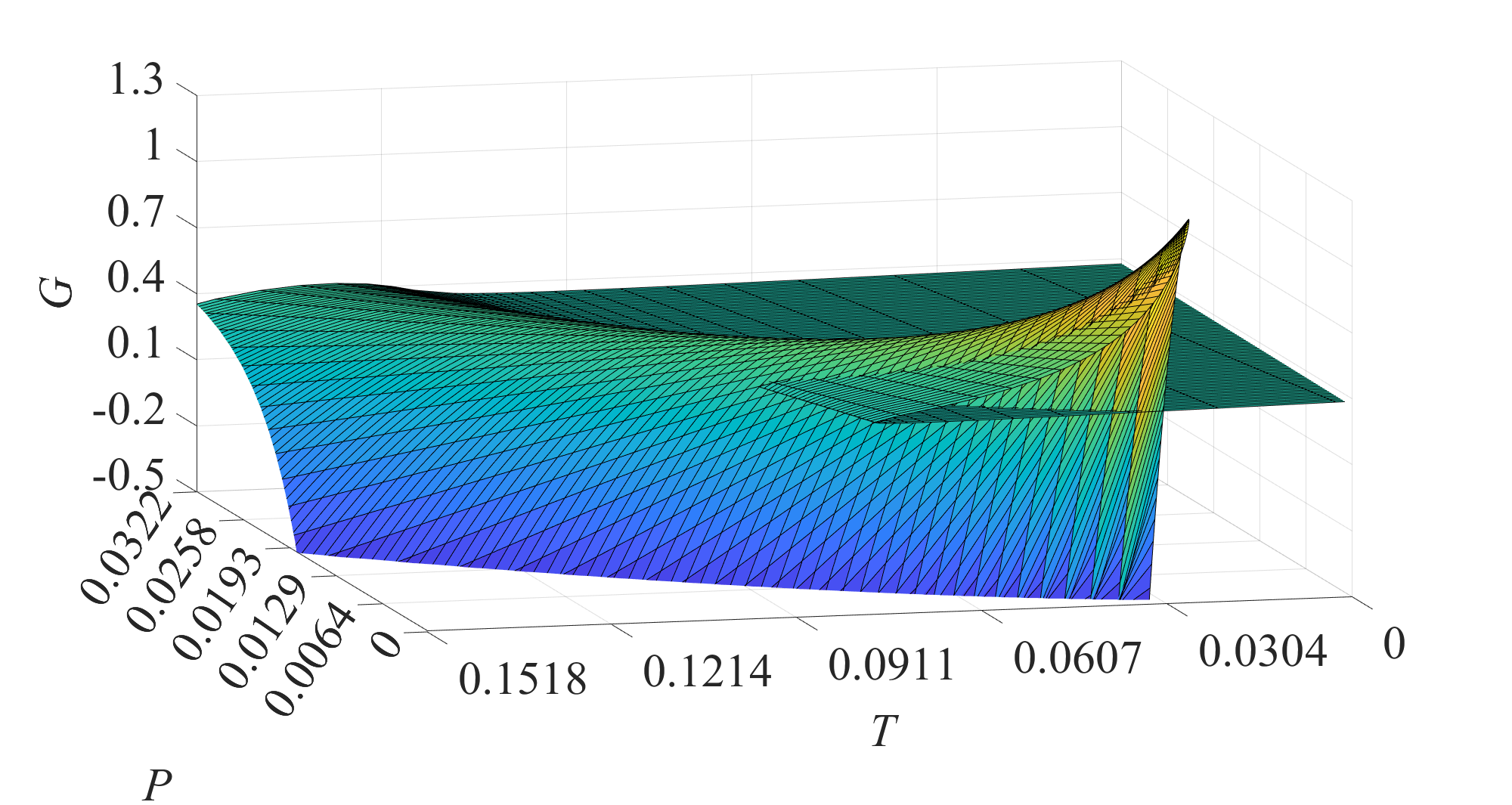}
\caption{}
\label{subfig:3Gibbs}
\end{subfigure}
\hfill
\begin{subfigure}[b]{0.45\textwidth}
\centering
\includegraphics[width=\textwidth]{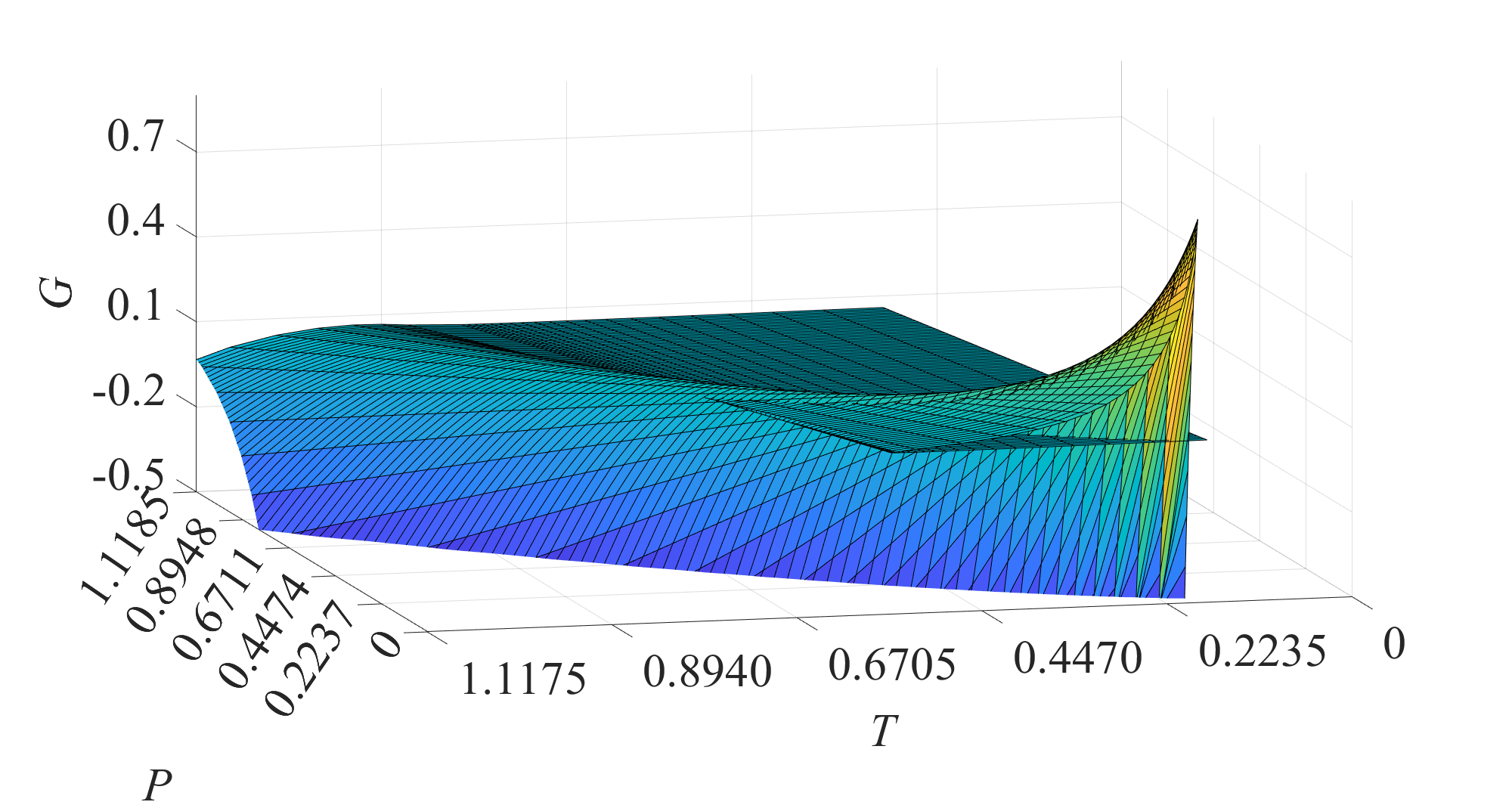}
\caption{}
\label{subfig:4Gibbs}
\end{subfigure}
\caption{Gibbs free energy surface diagrams. (a), (b), (c), and (d) correspond to the Case 1, 2, 3, and 4 respectively. These surface diagrams distinctly exhibit a swallowtail characteristic.}
\label{fig:Gibbs}
\end{figure}

\begin{figure}[htbp]
\centering
\begin{subfigure}[b]{0.45\textwidth}
\centering
\includegraphics[width=\textwidth]{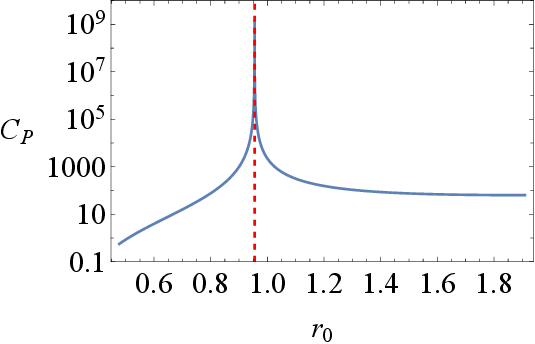}
\caption{}
\label{subfig:1CP}
\end{subfigure}
\hfill
\begin{subfigure}[b]{0.45\textwidth}
\centering
\includegraphics[width=\textwidth]{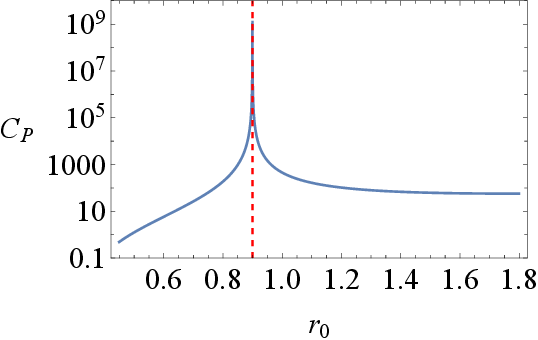}
\caption{}
\label{subfig:2CP}
\end{subfigure}

\begin{subfigure}[b]{0.45\textwidth}
\centering
\includegraphics[width=\textwidth]{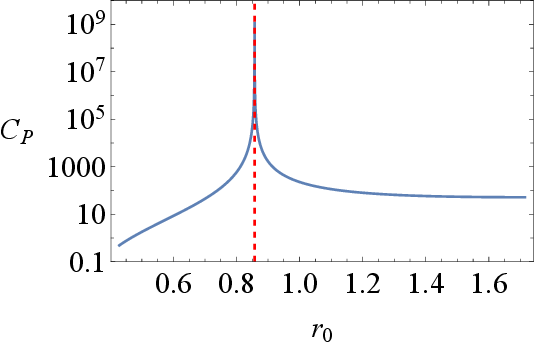}
\caption{}
\label{subfig:3CP}
\end{subfigure}
\hfill
\begin{subfigure}[b]{0.45\textwidth}
\centering
\includegraphics[width=\textwidth]{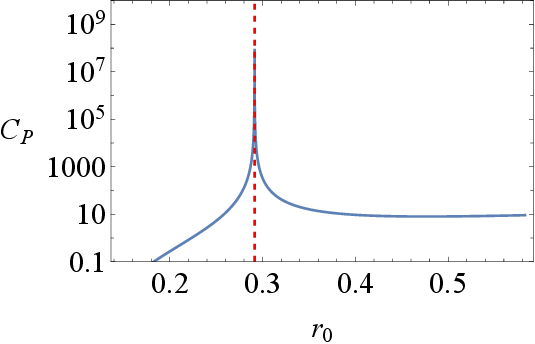}
\caption{}
\label{subfig:4CP}
\end{subfigure}
\caption{The isobaric capacity $C_P$ curves diagrams. (a), (b), (c), and (d) correspond to the Case 1, 2, 3, and 4 respectively. The blue curves respresent the $C_P$ curves, and the red dashed lines are dashed lines parallel to the $C_P$-axis at $r_0=r_{0c}$. It can be clearly seen that $C_P$ curves diverge to positive infinity at $r_0=r_{0c}$.}
\label{fig:CP}
\end{figure}

To verify whether the phase transition is a first-order transition of the system, we plot the Gibbs free energy $G(T,P)$ diagram and the isobaric heat capacity $C_P$ diagram based on equation \eqref{eq:36} as Fig.\ref{subfig:1Gibbs} and equation \eqref{eq:37} as FIG.\ref{subfig:1CP}. In FIG.\ref{subfig:1Gibbs}, the behavior of $G$-surface demonstrates a swallowtail characteristic. And in FIG.\ref{subfig:1CP}, the $C_P$ curve diverges at the critical point $r_0=r_{0c}$. These characteristics indicate the system exists a first-order phase transition.

\textbf{Case 2.}
When $r_0=0$, the expression for $h$ in equation \eqref{eq:25} gives
\begin{equation}
\label{eq:38}
\begin{split}
Q^2=& \frac{4qr_0^2 \sqrt{r_0^2+q^2}}{\gamma_2 \left( r_0^2+2q^2 \right)-\gamma_1 q \sqrt{r_0^2+q^2}} \\
& \times \left( \rule{0pt}{24pt} g^2 r_0^2+k \left( \frac{2r_0^2}{q^2} \log \left( 1+\frac{q^2}{r_0^2} \right)-\frac{r_0^2}{r_0^2+q^2} \right)+3\alpha r_0^2 \left( \frac{q}{\sqrt{r_0^2+q^2}}-\text{arcsinh} \left( \frac{q}{r_0} \right) \right) \right) \,.
\end{split}
\end{equation}
From equations \eqref{eq:23} and \eqref{eq:25}, combining the expressions of $T$, $h$, and $\sigma^2$ gives
\begin{equation}
\label{eq:39}
\begin{split}
& T=f_{T1} g^2 +f_{T2} \,,\\
& f_{T1}=\frac{\gamma _2 \left(q^2+r_0^2\right){}^2 \left(8 q^4+12 q^2 r_0^2+3 r_0^4\right)-4 \gamma _1 q \left(q^2+r_0^2\right){}^{7/2}}{4 \pi  r_0^5 \left(\gamma _2 \left(2 q^2+r_0^2\right)-\gamma _1 q \sqrt{q^2+r_0^2}\right)} \,,\\
& \begin{split}
f_{T2}= & \frac{\left( q^2+r_0^2 \right)^{3/2}}{4 \pi  q^2 r_0^5 \left(\gamma _1 q \left(q^2+r_0^2\right)-\gamma _2 \sqrt{q^2+r_0^2} \left(2 q^2+r_0^2\right)\right)} \\
& \times \left( \rule{0pt}{30pt} \gamma_1 q \sqrt{\left(q^2+r_0^2\right)} \left( \rule{0pt}{24pt} q^2 \left(3 \alpha  q \sqrt{q^2+r_0^2} \left(5 q^2+4 r_0^2\right)-2 k \left(4 q^2+3 r_0^2\right)\right) \right. \right. \\
& \mathrel{\hphantom{\times \left( \right.}} \quad \left.\rule{0pt}{24pt} +8 k \left(q^2+r_0^2\right){}^2 \log \left(\frac{q^2}{r_0^2}+1\right) - 12 \alpha  q^2 \left(q^2+r_0^2\right){}^2 \text{arcsinh} \left(\frac{q}{r_0}\right)  \right) \\
& \mathrel{\hphantom{\times \left( \right.}}+\gamma_2 \left( \rule{0pt}{24pt} \left(8 q^6+20 q^4 r_0^2+15 q^2 r_0^4+3 r_0^6\right) \left( 3 \alpha  q^2 \text{arcsinh} \left(\frac{q}{r_0}\right) -2 k \log \left(\frac{q^2}{r_0^2}+1\right) \right) \right.\\
& \mathrel{\hphantom{\times \left( \right.}} \quad \left. \rule{0pt}{30pt} \left.\rule{0pt}{24pt} +q^2 \left(k  \left(16 q^4+20 q^2 r_0^2+5 r_0^4\right)-3 \alpha  q \left(q^2+r_0^2\right){}^{3/2} \left(10 q^2+3 r_0^2\right)\right) \right) \right) \,.
\end{split}
\end{split}
\end{equation}
The thermodynamic pressure $P$ can be written as
\begin{equation}
\label{eq:40}
P=\frac{3}{8\pi} \frac{T-f_{T2}}{f_{T1}} \,.
\end{equation}
Considering $q << r_0$ and physical dimension, the trems involving $T$ in the expression for $P$ from equation \eqref{eq:40} can be approximated as
\begin{equation}
\label{eq:41}
\text{Press} \approx \frac{k\text{Temp}}{2 l_P^2 r_0} + \cdots \,.
\end{equation}
The specific volume $v$ can be identified with
\begin{equation}
\label{eq:42}
v=2l_P^2 r_0.
\end{equation}
The horizon radius $r_0$ also is associated with the fluid volume. Combining equations \eqref{eq:23}, \eqref{eq:39}, and $h(r_0)=0$, the Gibbs free energy $G$ of the system also can be written as
\begin{equation}
\label{eq:43}
G\left( T,P \right)=M-TS=\frac 12 \alpha q^3+\frac{\gamma_2 Q^2}{8q}-T \pi r_0^2 \,,
\end{equation}
where $Q^2$ is given by equation \eqref{eq:38} and $T$ is given by equation \eqref{eq:39}. The isobaric heat capacity $C_P$ and the isochoric heat capacity $C_V$ can be written as
\begin{equation}
\label{eq:44}
\begin{split}
& C_V=T \left. \frac{\partial S}{\partial T} \right|_V =0
\,,\\
&C_P=T \left. \frac{\partial S}{\partial T} \right|_P = \left. \frac{\partial S}{\partial r_0} \right|_P \left/ \left. \frac{\partial T}{\partial r_0} \right|_P \right.
=\left. \frac{\partial S}{\partial r_0} \right|_P \left/ \left(g^2 \left. \frac{\partial f_{T1}}{\partial r_0} \right|_P+ \left. \frac{\partial f_{T2}}{\partial r_0} \right|_P \right) \right. \,,
\end{split}
\end{equation}
where $f_{T1}$ and $f_{T2}$ are given by equation \eqref{eq:39} and $S$ is given by equation \eqref{eq:23}.

Substituting equation \eqref{eq:39} into equation \eqref{eq:31}, the critical point is $P_c \approx 0.019903334$, $r_{0c} \approx 0.899525553$, and $T_c \approx 0.097829295$. We also draw the $P-V$ diagram, $G(T,P)$-surface diagram, and $C_P$ curve diagram as the FIG.\ref{subfig:2PV}, FIG.\ref{subfig:2Gibbs}, and FIG.\ref{subfig:2CP} respectively. These three figures have identical characteristics with their counterparts in Case 1. Consequently, the static charged EMs black hole with $D=4$, $\Delta=1$, and $k_0=3$ also exhibits a first-order van der Waals-type phase transition.

\textbf{Case 3.}
When $r_0=0$, the expression for $h$ in equation \eqref{eq:26} gives
\begin{equation}
\label{eq:45}
\begin{split}
Q^2=& 12 r_0^2 \left(g^2 r_0^2+k-\frac{k q^4 \left(10 r_0^2+9 q^2\right)}{9 \left(r_0^2+q^2\right){}^3}-\frac{\alpha  r_0^2 q^3}{\left(r_0^2+q^2\right){}^{3/2}}\right) \\
& \times \frac{q \left(r_0^2+q^2\right){}^{3/2}}{\gamma _2 \left(3 r_0^4+12 q^2 r_0^2+8 q^4\right) -3 \gamma _1 q \left(r_0^2+q^2\right){}^{3/2}} \,.
\end{split}
\end{equation}
Similar to Case 2, from equations \eqref{eq:23} and \eqref{eq:26}, combining the expressions of $T$, $h$, and $\sigma^2$ yields
\begin{equation}
\label{eq:46}
\begin{split}
& T=f_{T1} g^2 +f_{T2} \,,\\
& f_{T1}= \frac{\gamma _2 \left(q^2+r_0^2\right){}^{3/2} \left(32 q^6+80 q^4 r_0^2+60 q^2 r_0^4+9 r_0^6\right)-12 \gamma _1 q \left(q^2+r_0^2\right){}^4}{4 \pi  r_0^4 \left(\gamma _2 \left(8 q^4+12 q^2 r_0^2+3 r_0^4\right)-3 \gamma _1 q \left(q^2+r_0^2\right){}^{3/2}\right)} \,,\\
&
\begin{split}
f_{T2}= & \frac{r_0 \left(\frac{q^2}{r_0^2}+1\right){}^{5/2}}{36 \pi } \left( \rule{0pt}{30pt} \frac{k \left(34 q^6+40 q^4 r_0^2\right)}{\left(q^2+r_0^2\right){}^4}-\frac{18 \alpha  q^3}{\left(q^2+r_0^2\right){}^{3/2}}+\frac{27 \alpha  q^3 r_0^2}{\left(q^2+r_0^2\right){}^{5/2}} \right. \\
& \quad +\left( 9 \alpha  q^3 \left(q^2+r_0^2\right){}^2-k \sqrt{q^2+r_0^2} \left(17 q^4+27 q^2 r_0^2+9 r_0^4\right) \right) \\
& \qquad \left. \rule{0pt}{30pt} \times \frac{\gamma _2 \left(16 q^6+40 q^4 r_0^2+30 q^2 r_0^4+3 r_0^6\right)-6 \gamma _1 q \left(q^2+r_0^2\right){}^{5/2}}{3 \gamma _1 q \left(q^2+r_0^2\right){}^6-\gamma _2 \left(q^2+r_0^2\right){}^{9/2} \left(8 q^4+12 q^2 r_0^2+3 r_0^4\right)} \right) \,.
\end{split}
\end{split}
\end{equation}
The expression of the thermodynamic pressure $P$ can also be written as equation \eqref{eq:40}. Considering $q << r_0$ and physical dimension, the trems involving $T$ in the expression for $P$ from equation \eqref{eq:33} can be approximated as
\begin{equation}
\label{eq:47}
\text{Press} \approx \frac{k \text{Temp}}{2 l_P^2 r_0}+\cdots \,.
\end{equation}
The specific volume $v$ can be written as equation \eqref{eq:42} and the horizon radius $r_0$ also is associated with the fluid volume. The form of the Gibbs free energy matches that given in equation \eqref{eq:43}, where $Q^2$ and $T$ are given by equations \eqref{eq:45} and \eqref{eq:46} respectively. And the isobaric heat capacity $C_P$ and the isochoric heat capacity $C_V$ are given by substituted $f_{T1}$ and $f_{T2}$ of equation \eqref{eq:46} and $S$ of equation \eqref{eq:23} into expression \eqref{eq:44}.

Taking equation \eqref{eq:46} into equation \eqref{eq:31}, the critical point is $P_c \approx 0.021483124$, $r_{0c} \approx 0.857180480$, and $T_c \approx 0.101188311$. We plot the $P-V$ diagram, $G(T,P)$-surface diagram, and $C_P$ curve diagram as the FIG.\ref{subfig:3PV}, FIG.\ref{subfig:3Gibbs}, and FIG.\ref{subfig:3CP} respectively. These three figures also have identical characteristics with their counterparts in Case 1 and Case 2. Therefore, the static charged EMs black hole with $D=4$, $\Delta=1$, and $k_0=5$ has a first-order van der Waals-type phase transition.

\textbf{Case 4.}
When $r_0=0$, the expression for $h$ in equation \eqref{eq:27} gives
\begin{equation}
\label{eq:48}
Q^2=\frac{4 q^2 r_0^4 \left(r_0^2 \left(2 \alpha +g^2\right)+k-2 \alpha  \sqrt{q^4+r_0^4}\right)}{\gamma _2 \sqrt{q^4+r_0^4}-\gamma _1 q^2} \,.
\end{equation}
From equations \eqref{eq:23} and \eqref{eq:27}, combining the expressions of $T$, $h$, and $\sigma^2$ gives
\begin{equation}
\label{eq:49}
\begin{split}
& T=f_{T1} g^2+f_{T2} \,,\\
& f_{T1}= \left(1+\frac{q^4}{r_0^4}\right) \frac{r_0 \left(\gamma _2 \left(3 q^4+2 r_0^4\right)-3 \gamma _1 q^2 \sqrt{q^4+r_0^4}\right)}{2 \pi \left( \gamma _2 \left(q^4+r_0^4\right)-\gamma _1 q^2 \sqrt{q^4+r_0^4}\right)} \,,\\
&\begin{split}
f_{T2}= \frac{1}{4 \pi} \left(1+\frac{q^4}{r_0^4}\right)& \left( \frac{4 \left(\sqrt{q^4+r_0^4} \left(k+3 \alpha  r_0^2\right)-\alpha  \left(2 q^4+3 r_0^4\right)\right)}{r_0 \sqrt{q^4+r_0^4}} \right. \\
& \quad \left. -\frac{2 \gamma _2 r_0^3 \left(k-2 \alpha  \sqrt{q^4+r_0^4}+2 \alpha  r_0^2\right)}{\gamma _2 \left(q^4+r_0^4\right)-\gamma _1 q^2 \sqrt{q^4+r_0^4}} \right) \,.
\end{split}
\end{split}
\end{equation}
Considering the expressions of $P$ in equation \eqref{eq:23} and equation \eqref{eq:49}, we obtain
\begin{equation}
\label{eq:50}
P=\frac{3}{4 \pi}g^2=\frac{3}{4\pi} \frac{T-f_{T2}}{f_{T1}}.
\end{equation}
Considering $q << r_0$ and physical dimension, the trems involving $T$ in the expression for $P$ in the above equation can be approximated as
\begin{equation}
\label{eq:51}
\text{Press}=\frac{3 k \text{Temp}}{4 l_P^2 r_0}+\cdots \,.
\end{equation}
The specific volume $v$ can be identified with
\begin{equation}
\label{eq:52}
v=\frac 43 l_P^2 r_0 \,.
\end{equation}
The horizon radius $r_0$ also is associated with the fluid volume. Combining equations \eqref{eq:23}, \eqref{eq:49}, and $h(r_0)=0$, the Gibbs free energy $G$ of the system can be written as
\begin{equation}
\label{eq:53}
G\left( T,P \right)=M-TS=\frac 32 \alpha q^4+\frac{3\gamma_2 Q^2}{8q^2}-2\pi r_0^3 T \,,
\end{equation}
where $Q^2$ and $T$ are given by equations \eqref{eq:47} and \eqref{eq:48} respectively. The isobaric heat capacity $C_P$ and the isochoric heat capacity $C_V$ are given by substituted $f_{T1}$ and $f_{T2}$ of equation \eqref{eq:49} and $S$ of equation \eqref{eq:23} into expression \eqref{eq:44}.

Taking equation \eqref{eq:49} into equation \eqref{eq:31}, the critical point is $P_c \approx 0.745638050$, $r_{0c} \approx 0.291352464$, and $T_c \approx 0.745010179$. We also plot the $P-V$ diagram, $G(T,P)$-surface diagram, and $C_P$ curve diagram as the FIG.\ref{subfig:4PV}, FIG.\ref{subfig:4Gibbs}, and FIG.\ref{subfig:4CP} respectively. These three figures also have identical characteristics with their counterparts in previous cases. Consequently, the static charged EMs black hole with $D=5$, $\Delta=2$, and $k_0=\frac 34$ exists a first-order van der Waals-type phase transition.

The first-order van der Waals-type phase transition is found in all four cases via analysis in this section. The positivity of the isobaric heat capacity in all four cases establishes the thermal stability of these black hole systems. Furthermore, it is revealed that the phase transition persists with decreasing scalar charge $q$. Conversely, when $q$ increases within a specific range, the phase transition remains observable. The critical point is not found by numerical methods when the scalar charge exceeds threshold values, which are approximately 2.0 (Case 1), 2.0 (Case 2), 1.9 (Case 3), and 2.2 (Case 4). This conclusively implies the absence of van der Waals-type criticality in the strongly scalar-charged regime. This absence presumably stemming from non-perturbative modifications of the gravitational system.

\subsection{Investigation on the Nonexistence of $P-V$ Phase Transitions}
In this subsection, we will firstly guarantee that the approximation of previous subsection remains valid in the strongly scalar-charged regime. For formal uniformity, the expressions $T$ and $P$ in Case 1 should be rewritten as the form of equations \eqref{eq:39} and \eqref{eq:40}. Comparing with the van der Waals equation \eqref{eq:11}, the specific volume $v$ for all four Cases can be identified with
\begin{equation}
\label{eq:54}
v=\frac 83 \pi l_P^2 f_{T1}\,.
\end{equation}
By substituting this volume for that of the four cases in the previous section, we find that the trend in the $P-V$ diagram and the phase transition point remain unchanged, i.e. $P_c,\,r_{0c}$, and $T_c$ remain unchanged. Furthermore, within the range of $r_0$ utilized in our phase transition research, $f_{T1}$ is monotonically increasing with $r_0$. Therefore, in the subsequent analysis, we can continue to treat $r_0$ as $v$.

\begin{figure}[htbp]
\centering
\begin{subfigure}[b]{0.45\textwidth}
\centering
\includegraphics[width=\textwidth]{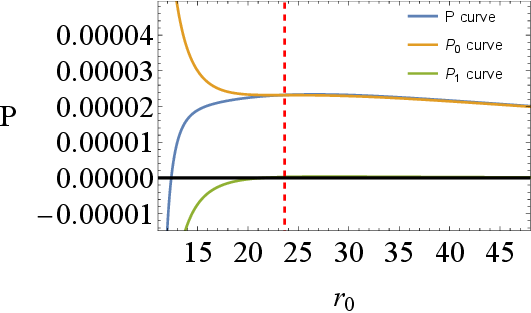}
\caption{}
\label{subfig:1PP0P1}
\end{subfigure}
\hfill
\begin{subfigure}[b]{0.45\textwidth}
\centering
\includegraphics[width=\textwidth]{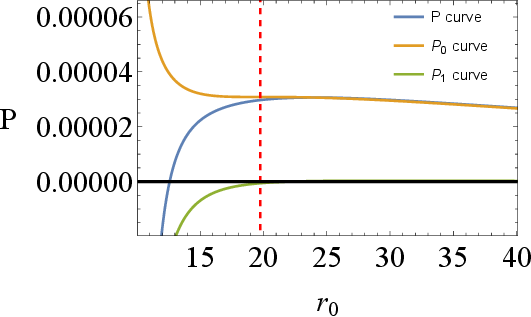}
\caption{}
\label{subfig:2PP0P1}
\end{subfigure}

\begin{subfigure}[b]{0.45\textwidth}
\centering
\includegraphics[width=\textwidth]{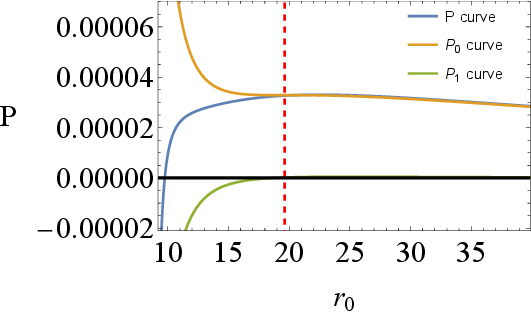}
\caption{}
\label{subfig:3PP0P1}
\end{subfigure}
\hfill
\begin{subfigure}[b]{0.45\textwidth}
\centering
\includegraphics[width=\textwidth]{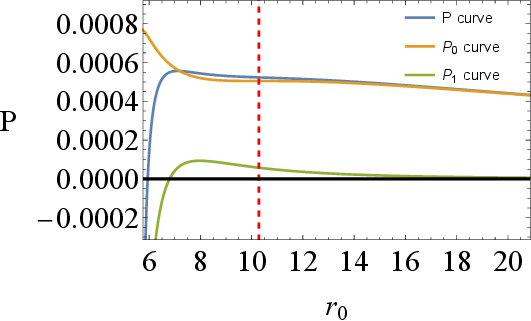}
\caption{}
\label{subfig:4PP0P1}
\end{subfigure}
\caption{The $P-v$, $P_0-v$, and $P_1-v$ diagrams for the all four cases. (a), (b), (c), and (d) correspond to the Case 1, 2, 3, and 4 respectively. In these diagrams, the blue, orange, and green curves correspond to the $P-V$, $P_0-v$, and $P_1-v$ curves respectively. $q=q_t+\delta q$ is slightly above $q_t$ in $P$ and $P_1$ curves and $q$ is slightly below $q_t$ in $P_0$ curves.}
\label{fig:PP0P1-v}
\end{figure}

\begin{figure}[htbp]
\centering
\begin{subfigure}[b]{0.45\textwidth}
\centering
\includegraphics[width=\textwidth]{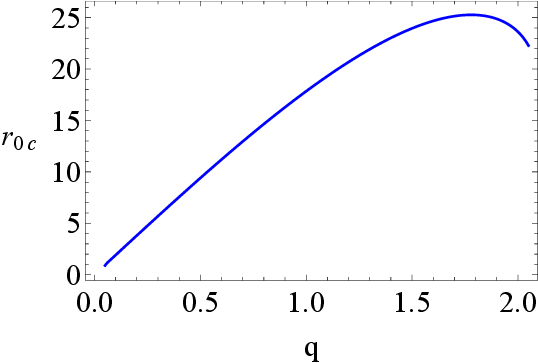}
\caption{}
\label{subfig:1qr0PointList}
\end{subfigure}
\hfill
\begin{subfigure}[b]{0.45\textwidth}
\centering
\includegraphics[width=\textwidth]{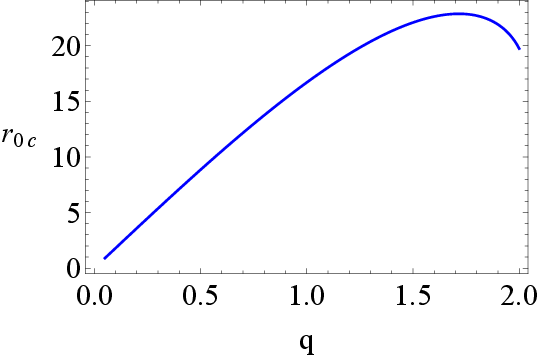}
\caption{}
\label{subfig:2qr0PointList}
\end{subfigure}

\begin{subfigure}[b]{0.45\textwidth}
\centering
\includegraphics[width=\textwidth]{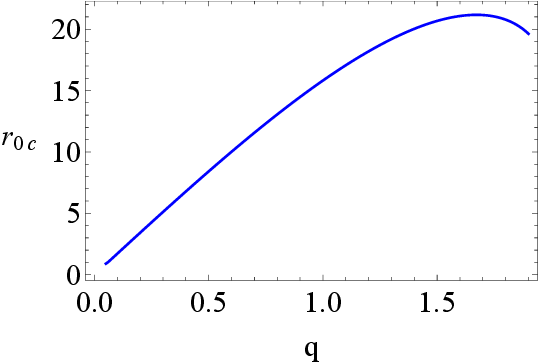}
\caption{}
\label{subfig:3qr0PointList}
\end{subfigure}
\hfill
\begin{subfigure}[b]{0.45\textwidth}
\centering
\includegraphics[width=\textwidth]{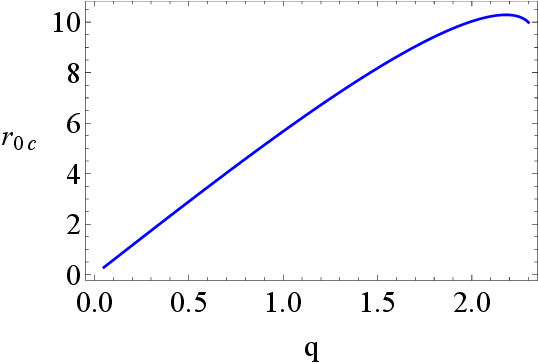}
\caption{}
\label{subfig:4qr0PointList}
\end{subfigure}
\caption{The $r_{0c}-q$ diagrams for the all four cases. (a), (b), (c), and (d) correspond to the Case 1, 2, 3, and 4 respectively. It can be seen from these diagrams that with increasing $q$, the trend of the $r_{0c}-q$ curves exhibit a transition point.}
\label{fig:r0c-q}
\end{figure}

\begin{figure}[htbp]
\centering
\begin{subfigure}[b]{0.45\textwidth}
\centering
\includegraphics[width=\textwidth]{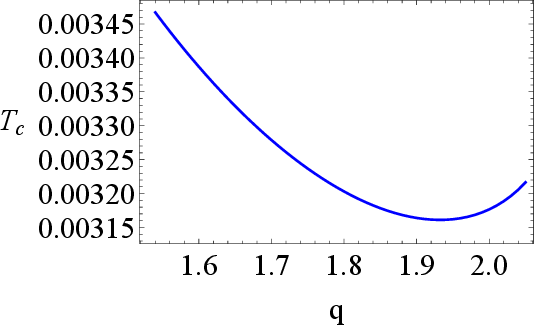}
\caption{}
\label{subfig:1qTPointList}
\end{subfigure}
\hfill
\begin{subfigure}[b]{0.45\textwidth}
\centering
\includegraphics[width=\textwidth]{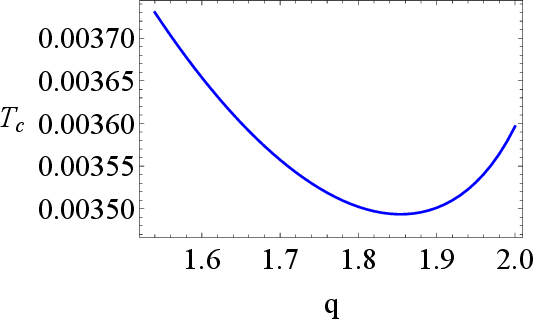}
\caption{}
\label{subfig:2qTPointList}
\end{subfigure}

\begin{subfigure}[b]{0.45\textwidth}
\centering
\includegraphics[width=\textwidth]{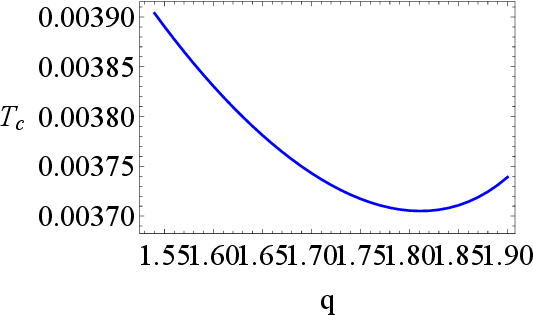}
\caption{}
\label{subfig:3qTPointList}
\end{subfigure}
\hfill
\begin{subfigure}[b]{0.45\textwidth}
\centering
\includegraphics[width=\textwidth]{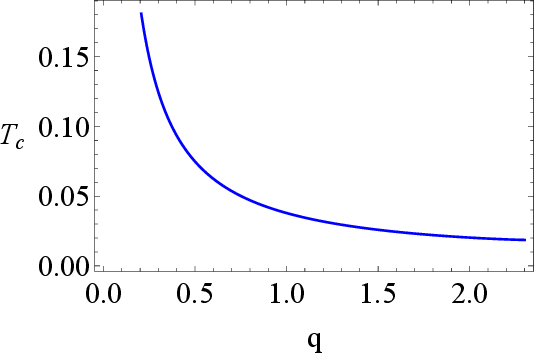}
\caption{}
\label{subfig:4qTPointList}
\end{subfigure}
\caption{The $T_c-q$ diagrams for the all four cases. (a), (b), (c), and (d) correspond to the Case 1, 2, 3, and 4 respectively. It can be seen from these diagrams that with increasing $q$, the trend of the $T_c-q$ curves exhibit a transition point for (a), (b), and (c), but curve does not exhibit a transition point for (d). To clearly demonstrate the transition point, we plot the portion with larger $q$, while the curve remains monotonically decreasing in the unplotted region of smaller $q$.}
\label{fig:Tc-q}
\end{figure}

In order to examine the behavior of phase transitions at $q_t$ (the threshold value of $q$), we perform the expansion of $P$ with respect to $q$ at $q$ slightly below $q_t$, while we denote the zeroth-order term as $P_0$ and the first-order term as $P_1$. In FIG.\ref{fig:PP0P1-v}, the $P-v$, $P_0-v$, and $P_1-v$ diagrams for all four cases are plotted, where $q=q_t+\delta q$ is slightly above $q_t$ in the $P$ and $P_1$ curves and $q$ is slightly below $q_t$ in the $P_0$ curves. We find that $P_0-v$ curves exhibit phase transitions, whereas $P-v$ curves do not. And $P_1-v$ curves diverge to negative infinity. Furthermore, we find that when $q$ is slightly below $q_t$, the trend of the $P_1-v$ curve aligns with FIG.\ref{fig:PP0P1-v};however, as $q$ continues to decrease, the trend reverses compared to the same figure. This trend variation is analyzed in the following. Before the trend variation, i.e., when $q_{tv} < q < q_t$ ($q_{tv}$ is the trend variation point), as $q$ increases, the divergence point of all four $P_1-v$ curves moves to the right, but it is still on the left side of the phase transition point. Therefore, we propose that when $q_{tv}<q<q_t$, as $q$ increases, the segment of $P$ on the left side of the phase transition point (corresponding to the segment of rapid divergence of $P_1$ to negative infinity) diminishes sharply. When $q$ reaches $q_t$, the phase transition point becomes the maximum point in the $P-v$ curve, leading to the disappearance of the phase transition. Furthermore, we also find that the critical volume shifts to the left as $q$ approaches $q_t$ from below, and shifts to the right when $q<q_{tv}$ for all cases. The shifting characteristic of critical temperature is the opposite of the critical volume in Case 1, 2, and 3, and in Case 4 $T_c$ decrease with rising $q$. The point where $r_{0c}(q)$ changes from monotone increasing to decreasing is marked as the volume transition point ($q_{rt}$), as $r_{0c}(q)$ remains a monotone increasing function over most interval of $q$. Analogously the point of $T_c(q)$ is marked as the temperature transition point ($q_{Tt}$) in Case 1, 2, and 3. It should be noted that $q_{tv} \neq q_{rt} \neq q_{Tt}$ in the general case and Case 4 has not temperature transition point. In FIG.\ref{fig:r0c-q} and FIG.\ref{fig:Tc-q}, we present $r_{0c}-q$ diagrams and $T_c-q$ diagrams for four Cases respectively. These figures provide clear evidence of the transition points in $r_{0c}-q$ curves in four cases and $T_c-q$ curves in Case 1, 2, and 3.

\section{Critical Phenomena} \label{sec:four}
In this section, we will study the critical behavior of the static charged EMs black hole near the critical point. The critical exponents are derived through the pressure series expansion.

To streamline the subsequent equations, we define the expression of $t$, $\phi$, and $p$ as
\begin{equation}
\label{eq:4-1}
t=\frac{T}{T_c}-1 \,,\quad \phi=\frac{v}{v_c}-1=\frac{f_{T1}\left( r_0 \right)}{f_{T1}\left( r_{0c} \right)}-1 \,,\quad p=\frac{P}{P_c} \,.
\end{equation}
The expression of $P$ is linear function of temperature, and satisfy the equation \eqref{eq:31} in the four exapmles. Then expanding around the critical point we approximate the pressure $p$ as
\begin{equation}
\label{eq:4-2}
\begin{split}
&p\left( t,\phi \right)=1 + c_1 t - c_2 t \phi - c_3 \phi^3 + o\left( t\phi^2,\phi^4 \right) \,, \\
&c_1=\left. \frac{\partial p}{\partial t} \right|_{\substack{t=0 \\ \phi=0}} \,,\quad c_2=-\left. \frac{\partial}{\partial \phi} \frac{\partial p}{\partial t} \right|_{\substack{t=0 \\ \phi=0}} \,,\quad c_3=-\left. \frac{\partial^3 p}{\partial \phi^3} \right|_{\substack{t=0 \\ \phi=0}} \,,
\end{split}
\end{equation}
where factly $c_1=c_2=\frac{3 T_c}{8 \pi P_c} \frac{1}{f_{T1}\left( r_{0c} \right)}$. We will justify $t \propto \phi^2$ rather than $t \propto \phi$ in equation \eqref{eq:4-4}. Differentiating the pressure $p$ for fixed $t<0$ we get
\begin{equation}
\label{eq:4-3}
\text{d}P=-P_c \left( c_2 t+3c_3 \phi^2 \right) \text{d}\phi\,.
\end{equation}
Considering the continuity of pressure and the Maxwell's equal area law $\oint v\text{d}P=0$ in the critical point, we get
\begin{equation}
\label{eq:4-4}
\begin{split}
& p=1+c_1 t-c_2 t\phi_s-c_3\phi_s^3=1+c_1 t-c_2 t\phi_l-c_3\phi_l^3 \,,\\
& 0=\int_{\phi_l}^{\phi_s} \phi\left( c_2 t+3c_3\phi^2 \right)\text{d}\phi \,.
\end{split}
\end{equation}
These equations have a unique non-trivial solution $\phi_l=-\phi_s=\sqrt{-c_2/c_3 \;t}$. Hence we get $t \propto \phi^2$.

The exponents $\alpha$, $\beta$, and $\gamma$ describe the behavior of the isochoric heat capacity $C_v$, order parameter $\eta$ and isothermal compressibility $\kappa_T$ respectively near the critical point and the exponent $\delta$ reflects the relation between $P$ and $v$ on the critical isotherm $T=T_c$. These exponents are defined by
\begin{equation}
\label{eq:4-5}
\begin{split}
& C_v=T \left. \frac{\partial S}{\partial T} \right|_v \propto |t|^{-\alpha} \,,\;\;\qquad \eta = v_l-v_s \propto |t|^{\beta} \,,\\
& \kappa_T=-\frac1v \left. \frac{\partial v}{\partial P} \right|_T \propto |t|^{-\gamma} \,,\qquad \left| P-P_c \right|_{T=T_c} \propto |v-v_c|^{\delta} \,.
\end{split}
\end{equation}
Considering equations \eqref{eq:23} and \eqref{eq:4-2}, the above equations can be written as
\begin{equation}
\label{4-6}
\begin{split}
& C_v=0 \,,\\
& \eta= v_c \left( \phi_l-\phi_s \right)=2v_c \sqrt{-\frac{c_2}{c_3}t} \,,\\
& \kappa_T=-\frac1v \left. \frac{\partial v}{\partial P} \right|_T=-\frac{1}{v_c\left( \phi+1 \right)} \frac{\partial v}{\partial \phi} \frac{\partial p}{\partial P} \left. \frac{\partial \phi}{\partial p} \right|_T = \frac{1}{c_2} \frac{1}{P_c} \frac 1t \,,\\
& \left| P-P_c \right|_{T=T_c}=\frac{1}{P_c}|p-1|=\frac{1}{P_c}\left| c_3 \phi^3 \right|=\left| \frac{c_3}{P_c v_c^3} \right| \left| v-v_c \right|^3 \,.
\end{split}
\end{equation}
If $c_2 \neq 0$ and $c_3 \neq 0$, these critical exponents can be straightforwardly deduced
\begin{equation}
\label{eq:4-7}
\alpha=0 \,,\quad \beta=\frac 12 \,,\quad \gamma=1 \,,\quad \delta=3 \,.
\end{equation}
We also calculate the isobaric heat capacity near the critical point as follows
\begin{equation}
\label{eq:4-8}
C_P-C_v=-T\left( \frac{\partial P}{\partial T} \right)_v^2 \left( \frac{\partial v}{\partial P} \right)_T=\frac{P_c v_c}{T_c}\frac{c_1^2}{2c_2}\frac{1}{-t} \,.
\end{equation}
This result proves that $C_P$ is indeed divergent with $c_2 \neq 0$ and $c_3 \neq 0$ at the critical point.

We will analyze the critical behavior of the black hole phase transition in four Cases at the critical point given in Section \ref{sec:three}. For all Cases, taking the critical values $P_c$, $r_{0c}$, and $T_c$ into expressions of $P$, the equation \eqref{eq:4-2} can be rewritten as
\begin{equation}
\label{eq:4-9}
\begin{split}
& \text{Case 1.}\quad p=1 + 2.98613448 t - 2.98613448 t\phi - 5.91214643 \phi^3 + o\left( t\phi^2,\phi^4 \right) \,, \\
& \text{Case 2.}\quad p=1 + 2.97728639 t - 2.97728639 t\phi - 6.08630854 \phi^3 + o\left( t\phi^2,\phi^4 \right) \,, \\
& \text{Case 3.}\quad p=1 + 3.00370561 t - 3.00370561 t\phi - 5.93641587 \phi^3 + o\left( t\phi^2,\phi^4 \right) \,, \\
& \text{Case 4.}\quad p=1 + 2.75211620 t - 2.75211620 t\phi - 6.68747743 \phi^3 + o\left( t\phi^2,\phi^4 \right) \,.
\end{split}
\end{equation}
We note $c_1,\,c_2,\,\text{and}\,c_3>0$ in all four Cases. This means that four critical exponents satisfy the equation \eqref{eq:4-7} and $C_P$ diverges to positive infinity. 

For 4-D and 5-D static charged EMs black holes, these critical exponents are independent of the dimensions of the black hole. The exponent $\alpha$ equals zero for black holes in arbitrary dimensions. The exponents $\beta,\,\gamma,\,\text{and}\,\delta$ also remain invariant for arbitrary dimensions on the condition that the series expansion coefficients satisfy $c_2 \neq 0$ and $c_3 \neq 0$. Compared with the van der Waals gas-liquid phase transition and the RN-AdS black hole phase transition, we find that all four critical exponents $\alpha,\,\beta,\,\gamma\,,\text{and}\,\delta$ of the static charged EMs black hole coincide with those of the van der Waals and RN-AdS systems. This result suggests a connection between the statistical properties of black holes and the mean field theory.

\section{ Conclusions} \label{sec:five}
This paper demonstrates the van der Waals-type phase transition and the critical behavior of the 4-D and 5-D static charged EMs black holes. Prior to initiating our investigation, we review the study of the RN-AdS black hole phase transition and the structure of the static charged EMs black hole solution. We also identify the cosmological constant and its conjugate quantity as the thermodynamic pressure and volume respectively. Taking 4-D and 5-D static charged EMs black holes as concrete cases, their $P-v$ diagrams, Gibbs free energy $G(T,P)$ diagrams and isobaric heat capacity $C_P$ diagrams are constructed. Subsequently, whether these diagrams have phase transition signatures is detailedly analyzed. We also study the critical behavior of the black hole near the critical point. The critical exponents are derived through the pressure series expansion.

The research in this paper reveals that in the 4-D case, the static charged EMs black hole has the following degenerations: (i) as the scalar charge approaches zero ($q \to 0$), the solution reduces to the RN-AdS solution; (ii) at the vanishing scalar charge ($q=0$), the solution reduces to the Schwarzschild-AdS solution. The $P-v$, $G(T,P)$, and $C_P$ diagrams for all four cases exhibit unambiguous phase transition signatures: (i) $P-v$ diagrams show the inflection point, (ii) $G(T,P)$ diagrams manifest the swallowtail characteristic, and (iii) $C_P$ curves diverge. The positive isobaric heat capacity guarantees the thermal stability of the static charged EMs black hole system. Furthermore, we find that the scalar charge has a ``threshold value" that determines whether the phase transition exists. The phase transition vanishes when the scalar charge exceeds the threshold value. In studying the disappearance of the phase transition at the threshold value, we identified a ``trend variation point" for the scalar charge: when $q$ is below this point, $\frac{\partial P}{\partial v}$ diverges to positive infinity; when $q$ exceeds this point, $\frac{\partial P}{\partial v}$ diverges to negative infinity. We also identified ``volume transition point" and ``temperature transition point" associated with the scalar charge. In all cases of the study, when $q$ is below the volume transition point, the critical volume increases with $q$ rising; when $q$ exceeds this point, the critical volume decreases with $q$ rising. The trend of the critical temperature is the opposite to that of the critical volume in some cases, while monotonically decreasing in others. This result implies that the small scalar charge can serve as a perturbation in gravitational systems, whereas large scalar charge cannot be treated as perturbations. This result can also be applied to the extended first law of thermodynamics incorporating scalar charges to study the thermodynamic effects of the scalar charge. It is found that the critical exponents $\alpha,\, \beta,\, \gamma$, and $\delta$ of the static charged EMs black hole coincide with those of the RN-AdS black hole and the van der Waals gas-liquid system, and are independent of the spacetime dimensionality of the black hole through the study of critical behavior. This suggests a connection between the statistical properties of black holes and the mean field theory.

\section{Acknowledgments}
We thank Prof Hong L$\ddot{\rm u}$ for helpful discussions.  This work is supported by the National Natural Science
Foundation of China (NSFC) under Grant nos. 12235019, 12275106, and Natural Science Foundation of Shandong
Province Nos.ZR2023MA014.

\end{document}